\documentclass[12pt]{article}
\pdfoutput=1
\usepackage{jheppub}
\usepackage{enumitem}
\usepackage{adjustbox}
\usepackage[T1]{fontenc} 
\usepackage{enumitem}  
\usepackage{empheq}
\usepackage{dsfont}
\usepackage[mathscr]{euscript}
\usepackage{dcolumn}
\usepackage{bm}
\usepackage{bbm}
\usepackage{slashed}
\usepackage{wrapfig}
\usepackage{mathtools}
\usepackage{tikz}
\usepackage{hyperref}

\usepackage{amsfonts,cases}
\usepackage{amsthm}

\newcommand{\dd}{\mathrm{d}}

\def\CF{{\cal F}}

\def\CM{{\cal M}}

\newcommand{\bes}[1]{\begin{equation} \begin{split} #1\end{split} \end{equation}}

\usepackage{tikz}
\usetikzlibrary{positioning}
\usetikzlibrary{arrows}
\usetikzlibrary{decorations.pathreplacing}
\usetikzlibrary{shapes.geometric}
\usetikzlibrary{calc}
\usetikzlibrary{decorations.pathmorphing}
\usetikzlibrary{shapes.misc}

\usetikzlibrary{positioning,trees,decorations.pathmorphing,decorations.markings,decorations.pathreplacing,calc,shapes,patterns,arrows,chains,arrows.meta,fit,fadings,decorations.markings,graphs,graphs.standard,quotes,plotmarks,calligraphy,decorations.pathreplacing}
\usetikzlibrary{intersections, calc}
\usetikzlibrary{shapes.geometric}
\usetikzlibrary{matrix, arrows.meta}
\usetikzlibrary{shapes.geometric}
\usetikzlibrary{decorations.pathmorphing}
\usetikzlibrary{shapes.misc}
\usetikzlibrary{arrows.meta}
\usetikzlibrary{angles, quotes}

\tikzset{flavour/.style={draw=none,minimum size=0.3mm,fill=white, regular polygon,regular polygon sides=4,draw}}
\tikzset{gaugeBig/.style={inner sep=1mm,draw=none,fill=white,minimum size=2mm,circle, draw}}
\tikzset{bd/.style={circle, draw=black, inner sep=0pt, fill=black, minimum size=2mm}}
\tikzset{wd/.style={circle, draw=black, inner sep=0pt, fill=white, minimum size=2mm}}
\tikzset{Dynkin/.style={circle, draw=black, inner sep=0pt, fill=white, minimum size=2mm}}
\tikzstyle{ligne}=[draw, very thick] 
\tikzstyle{gridline}=[draw, gray] 
\tikzset{gauge/.style={circle, draw,inner sep=2.5pt}}
\tikzset{gaugeo/.style={circle, draw,inner sep=2.5pt,fill=orange}}
\tikzset{gaugec/.style={circle, draw,inner sep=2.5pt,fill=cyan}}
\tikzset{gauger/.style={circle, draw,inner sep=2.5pt,fill=red}}
\tikzset{gaugeb/.style={circle, draw,inner sep=2.5pt,fill=blue}}
\tikzset{gaugeg/.style={circle, draw,inner sep=2.5pt,fill=green}}
\tikzset{gaugem/.style={circle, draw,inner sep=2.5pt,fill=magenta}}
\tikzset{hasse/.style={circle, fill,inner sep=2pt}}
\tikzset{shrinky/.style={circle, fill,inner sep=1pt}}
\tikzset{sized/.style={circle, draw, inner sep=1.5pt}}
\tikzset{seven/.style={circle, draw,inner sep=3pt}}

\tikzset{dotto/.style={circle, orange, draw,inner sep=1.5pt,fill=orange}}
\tikzset{dottp/.style={circle, purple, draw,inner sep=1.5pt,fill=purple}}
\tikzset{dottc/.style={circle, cyan, draw,inner sep=1.5pt,fill=cyan}}
\tikzset{dottr/.style={circle, red, draw,inner sep=1.5pt,fill=red}}
\tikzset{dottb/.style={circle, blue, draw,inner sep=1.5pt,fill=blue}}
\tikzset{dottg/.style={circle, green, draw,inner sep=1.5pt,fill=green}}
\tikzset{dottm/.style={circle, magenta, draw,inner sep=1.5pt,fill=magenta}}

\makeatletter
\gdef\@fpheader{}
\makeatother

\begin{document}

\setcounter{tocdepth}{2}

\title{New punctures for six-dimensional compactifications}

\author[1,2]{Fabio Apruzzi,}
\author[3,4]{Noppadol Mekareeya,}
\author[5]{Brandon Robinson,}
\author[3,6]{Alessandro Tomasiello}

    \affiliation[1]{Dipartimento di Fisica e Astronomia “Galileo Galilei”, Università di Padova,
    Via Marzolo 8, 35131 Padova, Italy}
    \affiliation[2]{INFN, Sezione di Padova Via Marzolo 8, 35131 Padova, Italy}
    \affiliation[3]{INFN, Sezione di Milano-Bicocca, Piazza della Scienza 3, I-20126 Milano, Italy}
    \affiliation[4]{Department of Physics, Faculty of Science, Chulalongkorn University, Phayathai Road, Pathumwan, Bangkok 10330, Thailand}
    \affiliation[5]{Institute of Physics, University of Amsterdam, Science Park 904, 1098 XH Amsterdam, Netherlands}
    \affiliation[6]{Dipartimento di Matematica, Università di Milano–Bicocca, Via Cozzi 55, 20126 Milano, Italy}

\emailAdd{fabio.apruzzi@pd.infn.it}
\emailAdd{n.mekareeya@gmail.com}
\emailAdd{b.j.robinson@uva.nl}
\emailAdd{alessandro.tomasiello@unimib.it}

\abstract{Six-dimensional super\-conformal field theories (SCFTs) give rise to four-dimensional (4d) ones when compactified on Riemann surfaces. In the $\mathcal{N}=(2,0)$ case, this yields the famous \emph{class S} family. For $\mathcal{N}=(1,0)$ theories that arise from linear unitary quivers, the holographic duals of the 4d theories are known in massive IIA supergravity, but only without punctures. Working in the probe approximation, we identify all possible BPS punctures in these models and characterize them by computing their defect Weyl anomalies. For class S, our results reproduce the known expressions in the appropriate limit. In the more general $\mathcal{N}=(1,0)$ case, they predict new 4d SCFTs and their large-$N$ anomaly coefficients.}
\maketitle

\section{Introduction}

In the study of super\-conformal field theories (SCFTs) in four dimensions, the so-called \emph{class S} is particularly intriguing. It is obtained by compactifying $\mathcal{N}=(2,0)$ theories in six dimensions on Riemann surfaces $\Sigma$ with or without punctures. It provides a rich playground of $\mathcal{N}=2$ theories with striking field-theoretic phenomena, including dualities \cite{Gaiotto:2009we} and holography \cite{Maldacena:2000mw,Gaiotto:2009gz}.

In the $A_{N-1}$ case, the basic building blocks are \emph{trinion} theories with $SU(N)^3$ flavor group, which can be gauged (fully or partially) to assemble a more general $\Sigma$.
We focus on \emph{regular} punctures on $\Sigma$, which are associated to ungauged $SU(N)$ groups, whose possible symmetry-breaking patterns are labeled by partitions of $N$. 
While trinions have no known Lagrangian realization, punctures often do admit explicit realizations as linear quivers.

One would expect to obtain a similar class of theories for any 6d $\mathcal{N}=(1,0)$ SCFT compactified on a Riemann surface. Some interesting results in this direction have already been obtained: for conformal matter theories \cite{Gaiotto:2015usa,Razamat:2022gpm,Kim:2018lfo}, for some special theories with no global symmetries \cite{Razamat:2018gro}, and/or in the presence of fluxes for the global symmetries \cite{Bah:2017gph,Kim:2017toz,Kim:2018bpg}.\footnote{In a separate line of research, one can also compactify an $\mathcal{N}=(2,0)$ theory to obtain $\mathcal{N}=1$ theories in four dimensions \cite{Maldacena:2000mw,Bah:2012dg,Gadde:2013fma,Bonelli:2013pva, Xie:2013rsa, Agarwal:2014rua, Giacomelli:2014rna, Bah:2015fwa}.}

In this paper, we make progress for the $\mathcal{N}=(1,0)$ theories whose effective theories consist of a linear chain of gauge groups $SU(r_i)$, $i=1,\,\ldots,\,N$. This is a vast class, whose massive IIA AdS$_7\times M_3$ holographic duals are known \cite{Apruzzi:2013yva,Apruzzi:2015wna,Cremonesi:2015bld}, as well as their AdS$_5\times \Sigma \times M_3$\footnote{As always in MN-type solutions, $M_3$ is actually fibered over $\Sigma$, and its metric is distorted with respect to the one appearing in the AdS$_7$ solutions.} compactifications 
\cite{Apruzzi:2015zna}, but only \emph{without} punctures. $M_3$ is obtained by fibering a round $\mathds{S}^2$ over an interval $[0,N]\ni z$; all fields are fully determined by a single piecewise-cubic function $\alpha(z)$, such that $\alpha''(i)=-81\pi^2 r_i$.  

Solutions with punctures are harder to obtain. Already for the ordinary $\mathcal{N}=2$ class S, they are known only locally around a single puncture, or on a sphere \cite{Gaiotto:2009gz}. A solution also exists for a certain choice of $\mathcal{N}=(1,0)$ theory and for the so-called simple type of puncture \cite{Bah:2017wxp}.

We avoid this difficulty by treating punctures as probes. First, as a warm-up, we look for BPS branes extended along an AdS$_5\times B_k$ in the AdS$_7\times M_3$ solutions (Secs.~\ref{sec:ads7} and \ref{sec:ads5}). These can be thought of as codimension-two defects $D_4$ for the 6d theory on $\mathbb{R}^6$, in a natural continuation of our earlier work on codimension-four defects \cite{Apruzzi:2024ark}.\footnote{Those defects might also play a role similar to the punctures in the present paper, by enriching the AdS$_3\times H_4/\Gamma \times M_3$ solutions in \cite[Sec.~5.2]{Passias:2015gya}.} But we then modify our computation to cover branes on AdS$_5\times\{\text{point}\}\times B_k$ in the AdS$_5\times \Sigma \times M_3$ solutions. These are defects for the 6d theory on $\mathbb{R}^4\times \Sigma$; in other words, punctures. 

We find two possible types of BPS configurations: a D4 brane, on a point $B_0$ on $M_3$ where $\alpha(z)$ attains a maximum; and a D6 brane with D4 charge, or in other words a D6/D4 bound state, on a $B_2\subset M_3$ such that $\alpha(z)\cos\theta=$ constant, with $\theta$ the polar angle on the internal $\mathds{S}^2$. These objects can combine over a point in $\Sigma$ to form more general punctures. (See Fig.~\ref{fig:sketch} below for a cartoon.) Recall that class S punctures arise from several stacks of M5-branes at $\mathbb{Z}_k$ singularities, which are indeed the uplifts of D6/D4 branes when the Romans mass vanishes. 

To obtain some quantitative checks on our results, we compute various Weyl anomaly coefficients (Sec.~\ref{sec:an}). $\langle T^\mu{}_\mu \rangle$ has a bulk contribution $\frac1{(4\pi)^{d/2}}E_d + c_I W_I$, where $W_I$ are various combinations of the Weyl tensor. There is also a defect contribution, which in $d=6$ is $\delta(D) \frac1{(4\pi)^2} (-a_{D} E_4 + \sum d_I J_I)$, where $E_4$ is the Euler density of the codimension-two defect $D$ and $J_I$ are 28 possible combinations---some preserving and others breaking defect parity---of the pullback of the ambient Weyl tensor, the second fundamental form, and of intrinsic and normal bundle curvatures \cite[(3.1)]{Chalabi:2021jud}. We holographically compute $a_D$ and one of the $d_I$ for both AdS$_7$ and AdS$_5\times \Sigma$. In the latter case we obtain for example
\begin{equation}\label{eq:aD4D6-intro}
	a_{\mathrm{D}4}= \frac{\alpha_\mathrm{max}}{48\pi^2} \, ,\qquad
	a_{\mathrm{D}6/\mathrm{D}4} = \frac1{96\pi^2} \int\alpha \,\dd z\,.
\end{equation}
Since we are working in the probe limit of AdS$_5$ branes, once we have computed one of the parity-even defect Weyl anomalies, the other 22 parity-even defect Weyl anomalies are fixed according to the ratios in \cite[Tab.~1]{Chalabi:2021jud} (see also \cite{Graham:2017bew}).

When we compactify the 6d theory on the Riemann surface $\Sigma$ with genus $g$, the Weyl anomaly will receive a bulk contribution from the integral on $\Sigma$ and a new localized contribution from the defects:\footnote{We do not turn on any flux for the abelian flavor symmetries of the 6d theory on $\Sigma$.}
\begin{equation}\label{eq:a4-a6-intro}    a_\mathrm{4d} =  (g-1) a_{\rm bulk} + \sum_i a_{\mathrm{D}_i}\,.
\end{equation}
$a_{\rm bulk}$ has been computed in general in terms of the 6d anomaly polynomial coefficients in \cite[(3.5)]{Bobev:2017uzs}. At large $N$ this becomes $a_\mathrm{bulk}=\frac{63}{512}a_\mathrm{6d}$ \cite[(3.6)]{Bobev:2017uzs}, with $a_\mathrm{6d}$ the leading-order Weyl anomaly of the 6d theory.
In this paper we will focus on the localized defect contributions at large $N$.
As reviewed in Sec.~\ref{sub:a-s}, this structure is indeed observed in class S theories \cite{Gaiotto:2009gz,Chacaltana:2010ks,Chacaltana:2012zy}; \eqref{eq:aD4D6-intro} agrees with the puncture contributions $a_{\mathrm{D}_i}$ whenever the probe approximation applies (Sec.~\ref{sub:match}). 

For compactifications of general $\mathcal{N}=(1,0)$ theories, currently we do not understand the 4d field theories well enough to check 
\eqref{eq:aD4D6-intro} independently. For class S$_k$ theories, namely the $\mathbb{Z}_k$ orbifolds of $\mathcal{N}=(2,0)$ on a Riemann surface,  we verify this relation in simple examples (Sec.~\ref{sub:Sk}). For more general $\mathcal{N}=(1,0)$ theories, \eqref{eq:aD4D6-intro} provides a large-$N$ prediction; in Sec.~\ref{sub:beyond} we spell this out in detail for some particular case. We hope a future field theory analysis will confirm these results.

\section{\texorpdfstring{Codimension-two probes in AdS$_7$}{Codimension-two probes in AdS7}} 
\label{sec:ads7}

In this section we will consider supersymmetric AdS$_7$ solutions of massive IIA, and BPS brane probes that are extended along a codimension-two subspace of AdS$_7$.

\subsection{\texorpdfstring{The AdS$_7$ solutions}{The AdS7 solutions}} 
\label{sub:ads7}

The AdS$_7 \times M_3$ background solutions to type IIA SUGRA constructed in \cite{Apruzzi:2013yva,Apruzzi:2015wna,Cremonesi:2015bld} is
\begin{align} \label{eq:ads7}
    \begin{split}
       \frac{1}{\pi\sqrt{2}}d s^2 &= 8 \sqrt{-\frac{\alpha}{\alpha''}}ds_{\mathrm{AdS}_7}^2 + \sqrt{-\frac{\alpha''}{\alpha}}\left(\dd z^2 +\frac{\alpha^2}{{\alpha'}^2 - 2\alpha\alpha''}\dd \Omega_2^2\right)\,,\\
       e^\phi &= 3^4(\sqrt{2}\pi)^\frac{5}{2}\frac{(-\alpha/\alpha'')^{\frac{3}{4}}}{\sqrt{{\alpha'}^2 -2\alpha\alpha''}}\,,\\
       B &= \pi\left(-z + \frac{\alpha\alpha'}{{\alpha'}^2 - 2\alpha\alpha''}\right)\text{vol}_{\mathds{S}^2}\,,\\
       F_2& = \left(\frac{\alpha''}{162\pi^2}+\frac{\pi F_0 \alpha\alpha'}{{\alpha'}^2 - 2 \alpha\alpha''}\right)\text{vol}_{\mathds{S}^2}\,.
    \end{split}
\end{align}
The warping function is $e^{2A}=8\sqrt{2}\pi\sqrt{-\alpha/\alpha''}$. $\dd \Omega_2^2$ is the $\mathds{S}^2$ metric, and ${\rm vol}_{\mathds{S}^2}$ its volume form. $\alpha$ satisfies
\begin{equation}\label{eq:a3}
    \alpha'''(z) = -162 \pi^3 F_0\,.
\end{equation}
Reality of the fields requires $\alpha\ge 0$, $\alpha''\le 0$. Flux quantization reduces to $-\alpha''(i)/81\pi^2$$\in \mathds{Z}$, $\forall i \in \mathds{Z}$, and to the domain of $z$ being $[0,N\in \mathds{Z}]$. 

When the Romans mass vanishes, these constraints single out
\begin{equation}\label{eq:F0=0}
	\alpha=\frac{81}2 k\pi^2 z (N-z)\,\qquad (F_0=0)\,;
\end{equation}
the corresponding solution uplifts to AdS$_7\times \mathds{S}^4/\mathds{Z}_k$ in $d=11$ supergravity. When $F_0\neq 0$, $\alpha''$ is piecewise linear; it can be parameterized as 
\begin{equation}\label{eq:a''-ri}
    -\frac{\alpha''(z)}{81 \pi^2} = r_i + (r_{i+1}-r_i) (z-i)\,.
\end{equation}
In each interval $z\in [i,i+1]$, $i\in \mathds{Z}$, we then have $r_i\in \mathds {Z}$. The loci where $F_0$ jumps are D8-branes. At the endpoints, several boundary conditions are possible; most notably, if $\alpha$ and $\alpha''$ have a single zero, the solution is regular. Other possibilities lead to the presence of D6-branes (a single zero for $\alpha$), O6-planes (a single zero for $\alpha''$), O8-planes (a double zero for $\alpha'$) \cite{Cremonesi:2015bld,Bah:2017wxp,Apruzzi:2017nck}.


\subsection{Supersymmetry} 
\label{sub:susy7}

We are interested in BPS branes in the solutions above. A quick way to find them is via pure forms \cite{Martucci:2005ht}, which were used to find these solutions in the first place \cite{Apruzzi:2013yva}. This technique is nicer in even dimensions; so we will initially consider AdS$_7\times M_3$ as a particular case of a Mink$_6\times M_4$ solution, with
\begin{equation}\label{eq:M4M3}
	\dd s^2_{M_4}= e^{2A}\frac{\dd \rho^2}{\rho^2} + \dd s^2_{M_3}\,;
\end{equation}
the function multiplying $\dd s^2_{\mathrm{Mink}_6}$ is $e^{2A_4}\equiv \rho^2 e^{2A}$. The supersymmetry equations now read \cite{Lust:2010by,Apruzzi:2013yva}
\begin{equation}\label{eq:pure64}
	\dd_H(e^{2A_4 - \phi} \Phi_0)=0 \, ,\qquad \dd_H(e^{4A_4 - \phi} \Phi_\alpha)= 0 \quad \alpha=1,2,3\, ,\qquad e^\phi F = 4 *_4\lambda (\dd A_4 \wedge \Phi_0)\,. 
\end{equation}
$F$ is the total RR flux in the $M_4$ directions; by definition, on a $k$-form we have $\lambda \beta_k = (-1)^{\lfloor k/2 \rfloor} \beta_k$. The pure forms are bilinears of the internal supersymmetry parameters:
\begin{equation}
	\Phi_0 = - 4\mathrm{Re} (\eta_+^1 \otimes \eta_-^{2\,\dagger}) \, ,\qquad
	\Phi_\alpha= 4\left(\mathrm{Re} (\eta_+^1 \otimes \eta_-^{2c\,\dagger}), \, \mathrm{Im} (\eta_+^1 \otimes \eta_-^{2c\,\dagger}), \mathrm{Im} (\eta_+^1 \otimes \eta_-^{2\,\dagger})\right) \,.
\end{equation}

The explicit expression of these pure forms for the solutions above were left a bit implicit in \cite{Apruzzi:2013yva}. They read
\begin{equation}
	\Phi_0 = \Psi_{0-} + e^A \frac{\dd \rho}\rho \wedge \Psi_{0+} 	\, ,\qquad
	\Phi_\alpha = \Psi_{\alpha-} + e^A \frac{\dd \rho}\rho \wedge \Psi_{\alpha+}
\end{equation}
with
\begin{equation}\label{eq:psi7}
\begin{split}
	&\Psi_{0-} = 4\pi e^{-A}\sqrt{1-X^2} \dd z + X \mathrm{vol}_3 
	\, ,\qquad
	\Psi_{0+}=
	 -X + \frac{e^{2A}}{16}(1-X^2)^{3/2}\mathrm{vol}_{\mathds{S}^2}\,,\\
	&\Psi_{\alpha-} = 4\pi X y_\alpha e^{-A} \dd z 
	+ \frac14 e^A\sqrt{1-X^2}\dd y_\alpha 
	- y_\alpha \sqrt{1-X^2}\mathrm{vol}_3 \, ,\\
	&\Psi_{\alpha+}= 
	 y_\alpha \sqrt{1-X^2} 
	+ y_\alpha \frac{e^{2A}}{16} X(1-X^2)\mathrm{vol}_{\mathds{S}^2}+\pi \sqrt{1-X^2} \dd z \wedge \epsilon_{\alpha \beta \gamma} y^\beta \dd y^\gamma\,.
\end{split}	
\end{equation}
Here $y_\alpha= (\sin \theta \cos \varphi, \sin \theta \sin \varphi, \cos \theta)$ are the usual coordinates embedding $\mathds{S}^2\hookrightarrow \mathds{R}^3$, and we defined 
\begin{equation}\label{eq:X}
	 X=\frac{\alpha'}{\sqrt{\alpha'{}^2- 2 \alpha \alpha''}}\,.
\end{equation}
The warping $A$ was given below (\ref{eq:ads7}). The index $\alpha$ is a triplet under the $\mathrm{SU}(2)_\mathrm{R}$ R-symmetry of the AdS$_7$ solutions.


\subsection{BPS probes} 
\label{sub:cal7}

We will now look for BPS branes whose worldvolume is of the form 
\begin{equation}
	M_{5+k}={\mathcal M}_5 \times B_k \subset \mathrm{AdS}_7 \times M_3\,.
\end{equation}
For now we will take ${\mathcal M}_5$ to be a flat $\mathds{R}^{1,4}$; $k$ can be only 0 or 2, for a D4 and a D6 respectively. We proceed by adapting \cite{Martucci:2005ht} to Mink$_6$ solutions, and specialize the resulting conditions to branes that are extended along the radial direction of AdS. The resulting BPS condition is equivalent to the brane being \emph{calibrated} by one of the $\Psi_{\alpha+}$, which without loss of generality we can take to be $\Psi_{3+}$:
\begin{equation}\label{eq:cal07}
	e^{-{\mathcal F}} \Psi_{3+}|= \sqrt{\det(g|+ {\mathcal F})} \dd \xi^1 \wedge \ldots \wedge \dd \xi^k\,;
\end{equation} 
This choice breaks the $\mathrm{SU}(2)_\mathrm{R}$ to a $\mathrm{U}(1)$.
Here $|$ denotes pull-back to $B_k$ (followed by projection to the $k$-form part, for a polyform); $\xi^{1\ldots k}$ are coordinates on $B_k$; and ${\mathcal F}= B + 2\pi l_s f$ as usual. An equivalent, and often more practical, condition is
\begin{equation}\label{eq:cal7}
	e^{-{\mathcal F}} \Psi_{0+}|=0 \, ,\qquad e^{-{\mathcal F}} \Psi_{\alpha+}|=0 \, \quad \alpha=1,2 
	 \,.
\end{equation}

The easiest case is $k=0$, a D4 placed at an internal point. In this case we only need to track the zero-form part of (\ref{eq:cal7}). From (\ref{eq:psi7}), the first in (\ref{eq:cal7}) reduces to $X=0$, which in turn gives
\begin{equation}\label{eq:alpha-max}
	\alpha'=0\,,
\end{equation}
whose solution we denote $\alpha_\mathrm{max}$. Indeed this is a maximum of $\alpha$, since $\alpha''\leq 0$ everywhere. 
The second in (\ref{eq:cal7}) then gives $\sin\theta=0$, fixing the point on the $\mathds{S}^2$ to be at either $\theta = 0$ or $\pi$. 

Next we consider D6-branes, which wrap a $B_2$ internal cycle. As a warm-up, we take this to span $z$ and the azimuthal angle $\varphi$  inside $\mathds{S}^2\subset M_3$, and we take the world-volume flux $f=0$. We note that $B$ vanishes when pulled back onto the $z,\,\varphi$ cycle, since it is proportional to $\text{vol}_{\mathds{S}^2}$. With this choice, the first in (\ref{eq:cal7}) is automatic, since $\Psi_{0+}$ only contains $\mathrm{vol}_{\mathds{S}^2}$. $\Psi_{1,2+}$ also contain terms proportional to $\sin \theta\cos \theta \dd \varphi$, which can be set to zero by taking $\theta=\pi/2$, so that the D6 sits on a great circle of the $\mathds{S}^2$. 

\begin{figure}
    \centering
    \includegraphics[width=8cm]{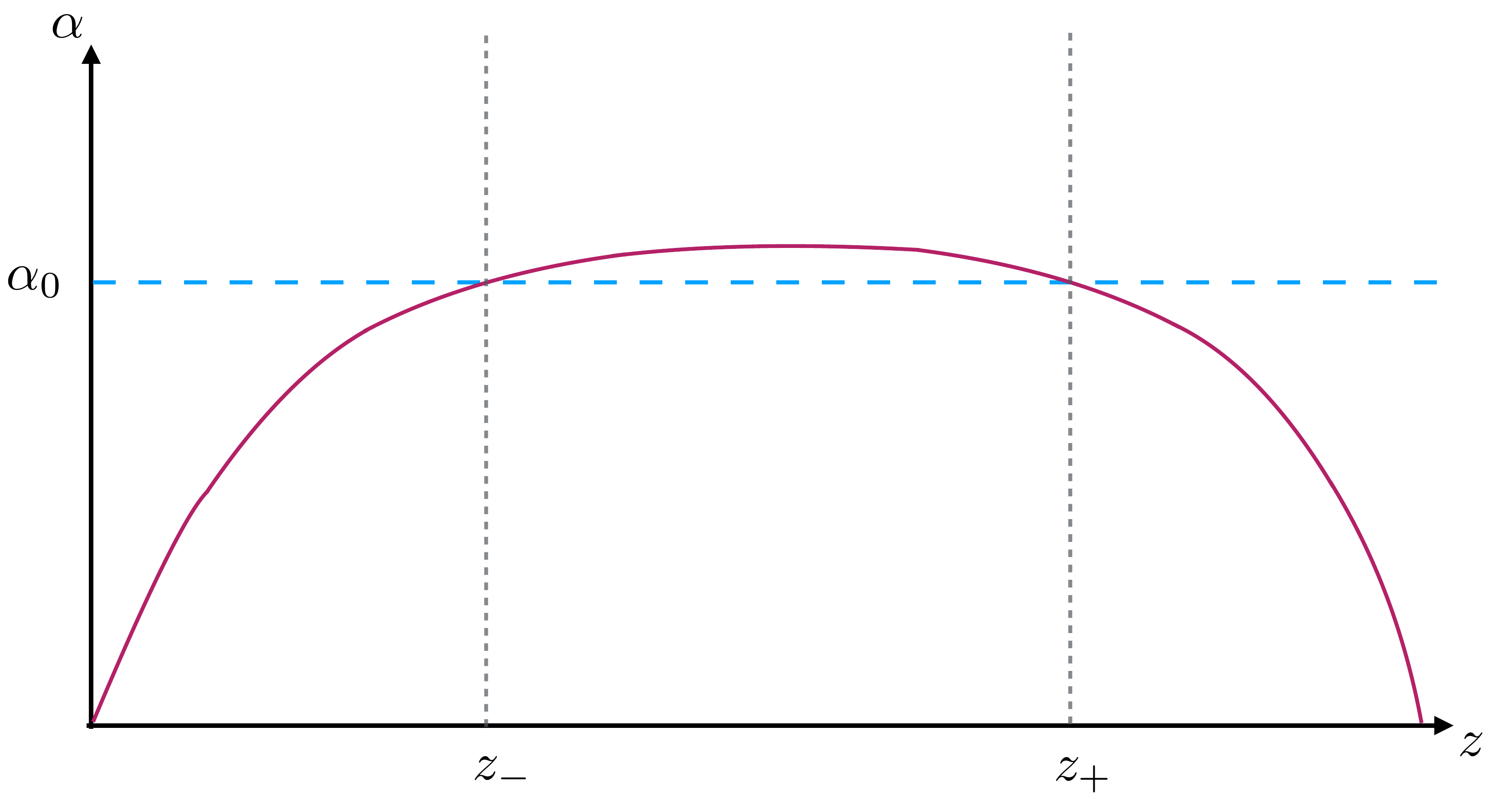}
	\quad
	\includegraphics[width=6cm]{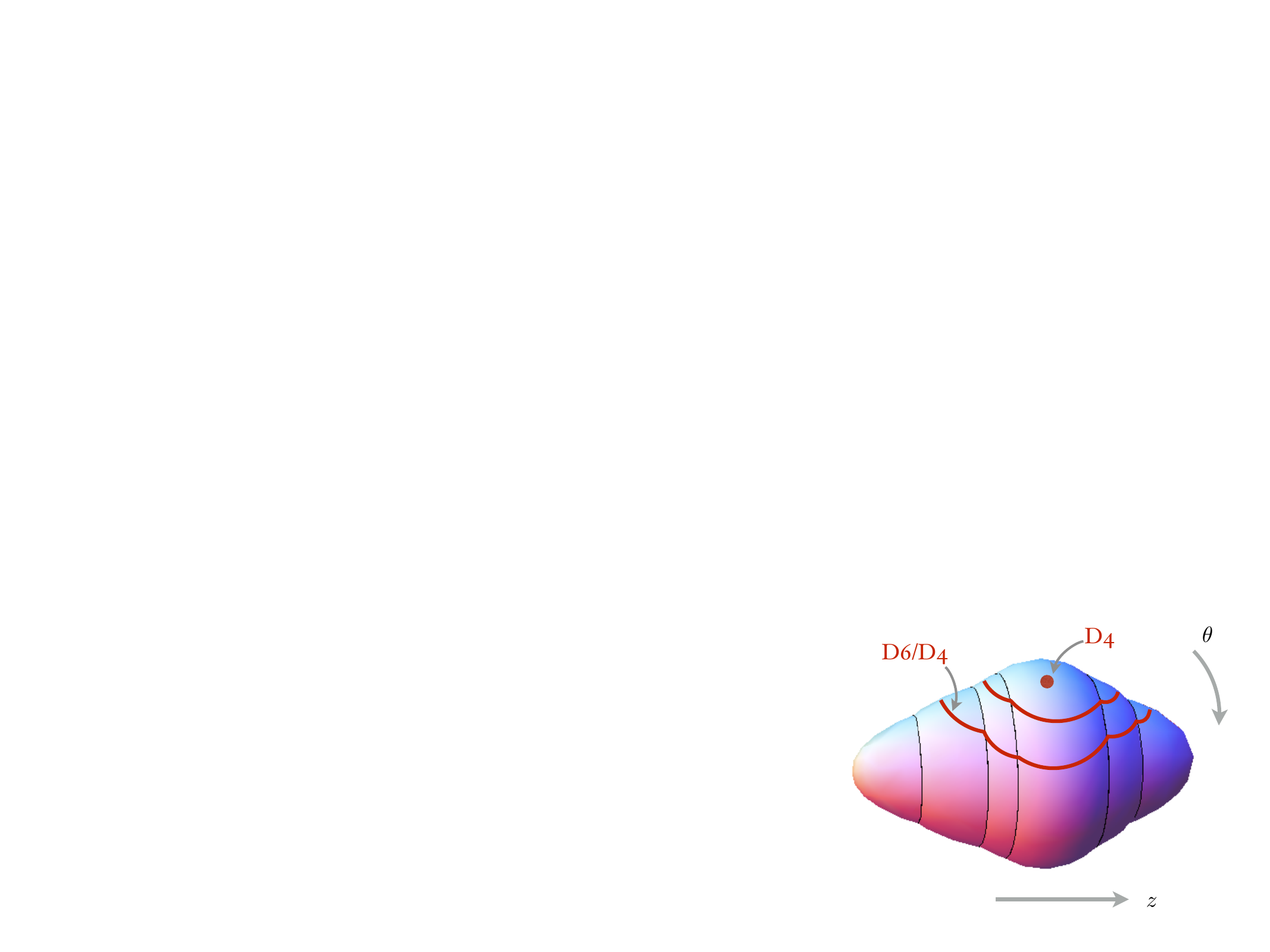}
    \caption{Left: The interval $[z_-,z_+]$ is defined as the region where $\alpha>\alpha_0$. The D6/D4 consists of a meridian of $\mathds{S}^2$ at $\cos\theta=\alpha_0/\alpha$, which degenerates to a point at $z=z_\pm$. Right: A two-dimensional cartoon of the internal space $M_3$ and of some brane probes. The $\mathds{S}^2$ is represented as a circle parameterized by $\theta$, suppressing the $\varphi$ direction. We show two examples of D6/D4 branes, and a single D4, which can be thought of as the limit $\alpha_0=\alpha_\mathrm{max}$. A general puncture is obtained by placing several of these objects over the same point in $\Sigma$.}
    \label{fig:sketch}
\end{figure}

There are also BPS probe D6/D4 bound states that interpolate between the pure D6 and D4 solutions above. The residual $\mathrm{U}(1)$ in (\ref{eq:cal7}) can be enforced as symmetry under rotations $\partial_\varphi$ on the $\mathds{S}^2$. So we take the $B_2$ extended along $\varphi$ and a combination of $\theta$ and $z$, parameterized by promoting
\begin{equation}
	\{\theta=\theta(z)\}\,.
\end{equation}
We write $B = b \mathrm{vol}_{\mathds{S}^2}$ (with $b$ known from (\ref{eq:ads7})), and now allow for a non-zero world-sheet flux $2\pi f = f_{z\varphi} \dd z\wedge \dd\varphi$.
(\ref{eq:cal7}) reduce to
\begin{subequations}\label{eq:D6D4} 
\begin{align}
	\label{eq:D6D4a} 
    e^{-\CF} \Psi_{0+}| &= (f_{z\varphi} - b \partial_z \cos\theta) X - \frac{e^{2A}}{16}(1-X^2)^{3/2}\partial_z\cos\theta = 0\\
    \label{eq:D6D4b} 
	e^{-\CF} \Psi_{1,2+}| &\propto (f_{z\varphi} - b \partial_z \cos\theta)+\frac{e^{2A}}{16}X\sqrt{1-X^2}\partial_z\cos\theta +\pi \cos\theta =0
\end{align}
\end{subequations}
Solving (\ref{eq:D6D4b})+$1/X$(\ref{eq:D6D4a}) gives
\begin{align}\label{eq:acos}
   \frac{\alpha^\prime}{\alpha}= -\frac{\partial_z\cos\theta}{\cos\theta}\qquad\Rightarrow\qquad \alpha \cos\theta = \alpha_0,\quad z\in[z_-,z_+]  
\end{align}
where $0\leq z_-\leq z_+\leq N$ and $\alpha_0$ is a constant.\footnote{If $\alpha$ were a radial variable, \eqref{eq:acos} would be the equation of a plane; this perhaps hints at the origin of this brane prior to back-reaction, which will be discussed qualitatively in Sec.~\ref{sub:brane}.}  The internal worldvolume $B_2$ of the brane is a fibration of the $\mathds{S}^1_\varphi$ over the interval $[z_-,z_+]$. The endpoints are the loci where $\alpha=\alpha_0$, and $\theta=0$ or $\pi$, so the $\mathds{S}^1_\varphi$ shrinks there. (Notice that $\alpha(z_-)= \alpha(z_+)= \alpha_0$.) All in all, $B_2$ has the topology of an $\mathds{S}^2\subset M_3$. All this is depicted in Fig.~\ref{fig:sketch}. 

Solving (\ref{eq:D6D4a}), the quantization of the 2-form flux becomes
\begin{align}\label{eq:qD6}
	f_{z\varphi}= - \alpha_0\pi \partial_z (z/\alpha) \quad \Rightarrow \quad
    \int_{B_2} f = -\alpha_0 \pi \frac{z}{\alpha}\Big|_{z_-}^{z_+} = -\pi(z_+-z_-) 
\end{align}
So, half the interval length $m\equiv \frac{z_+-z_-}2\in \mathbb{Z}$. This is the D4-charge. 

As $\alpha_0$ gets smaller, $z_\pm$ get closer to $N$ and $0$ respectively, and the two-cycle opens up. The D6-brane solution $\theta=\pi/2$ we saw earlier is a degenerate case, where $\alpha_0=0$ and hence $\int f=0$. So the D4 charge satisfies $m<N/2$.
On the other hand, as $\alpha_0\to\alpha_\mathrm{max}$, $z_\pm$ get closer together, and the two-cycle gets smaller, becoming more and more similar to a single D4-brane.

For later use we also evaluate the on-shell action. In general it is given by 
\begin{equation}\label{eq:SDp}
    S_{{\rm D}p} = T_p \int_{M_{p+1}}\dd^{p+1}x e^{-\phi}\sqrt{\det g| + \CF} - T_p\int_{M_{p+1}} e^{\mathcal F}\wedge C\, ,\qquad
	T_p = \frac{1}{2^p\pi^pl_s^{p+1}}\,.
\end{equation}
The first term, the DBI action, can be evaluated directly or using (\ref{eq:cal07}). The second term, the WZ action, vanishes, since the RR fields are purely internal, and their duals are of the form $\mathrm{vol}_{\mathrm{AdS}_7}\wedge$ internal.

For the D4 we obtain
\begin{equation}\label{eq:SD4-7}
	S_{\mathrm{D}4}= T_4 \mathrm{Vol}({\mathcal M}_5) e^{5A-\phi}|_{\mathrm{D}4}= \mathrm{Vol}({\mathcal M}_5)\frac{16}{81\pi^4} \alpha_\mathrm{max}\,.
\end{equation}
For the D6/D4,
\begin{equation}\label{eq:SD6-7}
	S_{\mathrm{D}6/\mathrm{D}4}= T_6 \mathrm{Vol}({\mathcal M}_5) \int e^{5A-\phi} e^{-{\mathcal F}} \Psi_{3+}= \mathrm{Vol}({\mathcal M}_5)\frac8{81\pi^4} \int_{z_-}^{z_+} \alpha \dd z \,.
\end{equation}
The factor $\mathrm{Vol}({\mathcal M}_5)$ diverges, because of contributions near the conformal boundary. In the next subsection we will carefully treat the holographic renormalization of this divergence.

For the massless case (\ref{eq:F0=0}), we obtain
\begin{equation}\label{eq:SD-F0-7}
	S_{\mathrm{D}4}= \mathrm{Vol}({\mathcal M}_5)\frac2{\pi^2}kN^2 \, ,\qquad
	S_{\mathrm{D}6/\mathrm{D}4}= \mathrm{Vol}({\mathcal M}_5) \frac{2}{\pi^2}k m \left(N^2-\frac43 m^2\right)\,.
\end{equation}


\subsection{Brane interpretation} 
\label{sub:brane}

The AdS$_7$ solutions are thought to arise from a near-horizon limit of a brane configuration \cite{Gaiotto:2014lca}. We can interpret our probe branes in this section as arising from additional branes. We summarize the result in Table \ref{tab:brane-diagram}.

\begin{table}[ht]
    \centering
    \label{tab:brane-diagram} 
    \begin{tabular}{|c||cccccccccc|}
        \cline{2-11} 
        \multicolumn{1}{c|}{} & $0$ & $1$ & $2$ & $3$ & $4$ & $5$ & $6$ & $7$ & $8$ & $9$ \\
        \hline
        \text{NS5} & $\times$ & $\times$ & $\times$ & $\times$ & $\times$ & $\times$ & $-$ & $-$ & $-$ & $-$ \\
        \text{D6}  & $\times$ & $\times$ & $\times$ & $\times$ & $\times$ & $\times$ & $\times$ & $-$ & $-$ & $-$ \\
        \text{D8}  & $\times$ & $\times$ & $\times$ & $\times$ & $\times$ & $\times$ & $-$ & $\times$ & $\times$ & $\times$ \\
        \hline
        \text{def. D4}  & $\times$ & $-$ & $-$ & $\times$ & $\times$ & $\times$ & $-$ & $-$ & $-$ & $\times$ \\
        \text{def. D6}  & $\times$ & $-$ & $-$ & $\times$ & $\times$ & $\times$ & $\times$ & $\times$ & $\times$ & $-$ \\
        \hline
    \end{tabular}
    \caption{Brane diagram engineering the 6d SCFT and its defects.} 
\end{table}

T-dualizing along some of the field theory directions yields similar defects in\allowbreak{} SCFTs in $d\neq6$. 
For example, with directions 456 we obtain the famous Hanany--Witten brane engineering with D3, D5 and NS5 \cite{Hanany:1996ie}, along with defect D1 and D3 and possibly a D3/D1 bound state. 
The D1 is the vortex in \cite{Hanany:2003hp}. 
The defect D3 can be thought of as a pole for a complex field along the D3s making up the SCFT. 
The latter are now extended along 0126; if one groups directions 12 and 78 into two complex coordinates $z$ and $w$ respectively, this configuration would be described by a locus $\{zw=1\}$. It might be interesting to investigate these defects further.

The D6/D4 bound state of the previous subsection is obtained by fusing the defect D4 and D6 above in a funnel, which can be described as a solution of the Nahm equation on the D4 with a pole, or as a monopole on the D6. This is similar to how the non-defect D6 and D8 fuse together in a Nahm pole on the D6. The D8/D6 bound states in the AdS$_7$ solution are thought to be remnants of such Nahm poles after the near-horizon limit \cite{Gaiotto:2014lca}. 

Performing a T-duality along direction 7 gives rise to a different realization in IIB. In this picture, both the non-defect D6 and the D8 become D7s; the resulting D7 intersection is then a particular case of the F-theory engineering of 6d SCFTs initiated in \cite{Heckman:2013pva}. The defect D4 and D6 both become D5 in this picture. The Nahm poles should now turn into Hitchin poles. From the point of view of the D7s, a pole in the 67 direction describes the data of the 6d theory, while a pole in the 12 direction would describe the defect. The relevant BPS equation should be an analogue of the Hitchin equation with two commuting scalars, which is known to arise on brane systems. A similar perspective was given for punctures in class S$_k$ theories in \cite{Heckman:2016xdl}. 



\section{\texorpdfstring{Codimension-two probes in AdS$_5$}{Codimension-two probes in AdS5}} 
\label{sec:ads5}

We will now consider massive IIA AdS$_5\times M_5$ solutions where the internal space is a fibration over a Riemann surface: 
\begin{equation}
	M_3 \hookrightarrow M_5 \to \Sigma\,.
\end{equation}
These should represent the gravity duals of the compactifications on $\Sigma$ of the ${\mathcal N}=(1,0)$ 6d SCFTs dual to the AdS$_7$ solutions (\ref{eq:ads7}) \cite{Apruzzi:2015wna}. They are hence similar to the Maldacena--N\'u\~nez solutions \cite{Maldacena:2000mw}, whose dual are the Riemann surface compactifications of the ${\mathcal N}=(2,0)$ theories. Within these solutions, we want to consider defects that are extended along AdS$_5$, and are point-like in $\Sigma$. Eventually we will interpret these as a lower-supersymmetry counterpart of the punctures of \cite{Gaiotto:2009gz}.

\subsection{\texorpdfstring{AdS$_5$ compactifications of the AdS$_7$ solutions}{AdS5 compactifications of the AdS7 solutions}} 
\label{sub:ads5}

The AdS$_5$ solutions we need were found in \cite{Apruzzi:2015wna}. They read:
\begin{align} \label{eq:ads5}
    \begin{split}
       \dd s^2 &= 3\sqrt{6}\pi \sqrt{-\frac{\alpha}{\alpha''}}\left(\dd s_{\mathrm{AdS}_5}^2 + \dd s^2_\Sigma\right)+ \pi\sqrt{-\frac32\frac{\alpha''}{\alpha}}\left(\dd z^2 +\frac{\alpha^2}{{\alpha'}^2 - \frac32 \alpha\alpha''}\widetilde{\dd \Omega}_2^2\right)\,,\\
       e^\phi &= 2^{3/4} 3^{17/4}\pi^{5/2}
	   \frac{(-\alpha/\alpha'')^{\frac{3}{4}}}{\sqrt{{\alpha'}^2 -\frac32\alpha\alpha''}}\,,\\
       B &= \pi\left(-z + \frac{\alpha\alpha'}{{\alpha'}^2 - \frac32\alpha\alpha''}\right)\widetilde{\text{vol}}_{\mathds{S}^2}-3\pi z \cos\theta \mathrm{vol}_\Sigma\,,\\
       F_2& = \left(\frac{\alpha''}{162\pi^2}+\frac{\pi F_0 \alpha\alpha'}{{\alpha'}^2 - \frac32 \alpha\alpha''}\right)\widetilde{\text{vol}}_{\mathds{S}^2}+\frac{\alpha''}{54\pi^2}\cos \theta \mathrm{vol}_\Sigma \,,\\
	   F_4& = -\frac1{54\pi}\sin \theta \alpha'' D \psi \wedge \mathrm{vol}_\Sigma \wedge\left( \frac{\alpha \alpha'}{\alpha'{}^2 -\frac32 \alpha \alpha''}\cos \theta  \dd\theta + \sin \theta \dd z \right)\,.
    \end{split}
\end{align}
$\Sigma$ is a Riemann surface of scalar curvature $=-3$, and hence of genus $g\ge 2$. The tilde means fibration over $\Sigma$: $\widetilde{\dd \Omega}_2^2= \dd \theta^2 + \sin^2 \theta D \psi^2$, $\widetilde{\text{vol}}_{\mathds{S}^2}= \sin \theta \dd \theta \wedge D \psi$, where $D \psi = \dd \psi - a$, $\dd a= -3 \mathrm{vol}_\Sigma$. We note that the warping function multiplying AdS$_5$ is
\begin{equation}
	e^{2A_5}= 3\pi\sqrt{-6\frac \alpha{\alpha''}}\,.
\end{equation}

As we mentioned, these solutions are similar to the Maldacena--N\'u\~nez solutions. More precisely, for $F_0=0$ (\ref{eq:ads5}) become the ${\mathcal N}=1$ MN solutions \cite[Sec.~4.2]{Maldacena:2000mw}.\footnote{In terms of the expression of the solution in \cite[(5.15)]{Gauntlett:2004zh}, the coordinate change is $z=N \sin^2(\alpha/2)$ where $\alpha$ is now an angle used in that reference.} 

For later use, we give here the \emph{external} RR potentials, namely those of the type $\mathrm{vol}_{\mathrm{AdS}_5}\wedge (\ldots)$. In general in IIA we have $e^{-B} F=F_0 + \dd (e^{-B} C) $, $* \lambda F= F$. For example $(e^{-B}F)_6= F_6 - B \wedge F_4 +\frac12 B^2 F_0$, but the last two terms are purely internal, so the external five-form potential satisfies $F_6=*F_4 = \dd C_5^\mathrm{ext}$. For the seven-form potential, we want $F_8-B \wedge F_6 = - * F_2 - B \wedge * F_4 = \dd(e^{-B} C)^\mathrm{ext}_7$. A possible expression for these potentials is
\begin{equation}\label{eq:Cext}
	\begin{split}
	C_5^\mathrm{ext}=&\frac \alpha 3 \cos \theta \mathrm{vol}_{\mathrm{AdS}_5} \, ,\\
		 (e^{-B} C)^\mathrm{ext}_7 =& 
		 \pi\Big[\left(6\frac{\alpha \alpha'}{\alpha''}+ z\alpha \cos^2 \theta- \hat\alpha(15+\cos^2 \theta)\right) \mathrm{vol}_\Sigma \\
		 &\quad +\frac16(z \alpha-2 \hat \alpha) \sin(2 \theta)\dd\theta \wedge D \psi \Big]\wedge \mathrm{vol}_{\mathrm{AdS}_5}\,.
	\end{split}
\end{equation}
Here $\hat \alpha'= \alpha$.\footnote{Between any two D8-brane stacks one can use $(\frac{2 \alpha \alpha''- \alpha'{}^2}{-324 \pi^3 F_0 })'= \alpha$.} One can check that these forms are regular at the poles of the $\mathds{S}^2$, and at $z=0$, $N$ if the solution is regular there. (Recall from Sec.~\ref{sub:ads7} that this corresponds to a single zero for $\alpha$ and $\alpha''$.)


\subsection{Supersymmetry} 
\label{sub:susy5}

Again we describe the supersymmetric data of the solutions by recasting it as a Mink$_4\times M_6$ solution, similar to (\ref{eq:M4M3}): $\dd s^2_{M_6}= e^{2A}\dd \rho^2/\rho^2 + \dd s^2_{M_5}$, and with the function multiplying $\dd s^2_{\mathrm{Mink}_4}$ being $e^{2A_6}\equiv \rho^2 e^{2A}$. The supersymmetry equations now read \cite{Grana:2005sn}
\begin{equation}
	\dd_H(e^{2A_6-\phi} \mathrm{Re} \Phi_-)= 0 \, ,\qquad
	\dd_H(e^{3A_6-\phi}\Phi_+)=0 \, ,\qquad \dd_H(e^{4A_6-\phi}\mathrm{Im} \Phi_-)=e^{4A_6} *_6 \lambda F\,.
\end{equation}
$F$ is now the total RR field along $M_6$.

The pure forms $\Phi_\pm$ can be extracted with some work from \cite{Apruzzi:2015wna}:
\begin{equation}
	\Phi_+ = \eta_+^1 \otimes \eta_+^{2\,\dagger}= (\phi_1 + i \phi_2) \wedge E \, ,\qquad
	\Phi_- = \eta_+^1 \otimes \eta_-^{2\,\dagger}= - (\phi_0 + i \phi_3) \wedge \exp[E \wedge \bar E/2]\,;
\end{equation}
$E$ is a $(1,0)$-form such that $E \bar E= \dd s^2_\Sigma$. As in Sec.~\ref{sub:susy7}, we can decompose
\begin{equation}
	\phi_0 = \psi_{0-} + e^A \frac{\dd \rho}\rho \wedge \psi_{0+} 	\, ,\qquad
	\phi_\alpha = \psi_{\alpha-} + e^A \frac{\dd \rho}\rho \wedge \psi_{\alpha+}\,,
\end{equation}
but now we have
\begin{equation}\label{eq:psi5}
\begin{split}
	&\psi_{0-} = 3\pi e^{-A_5}\sqrt{1-X_5^2} \dd z + X_5 \widetilde{\mathrm{vol}}_3 
	\, ,\qquad
	\psi_{0+}=
	 -X_5 + \frac{e^{2A_5}}{9}(1-X_5^2)^{3/2}\widetilde{\mathrm{vol}}_{\mathds{S}^2}\,,\\
	&\psi_{\alpha-} = 3\pi X_5 \hat y_\alpha e^{-A_5} \dd z 
	+ \frac13 e^{A_5}\sqrt{1-X_5^2}\dd\hat y_\alpha 
	- y_\alpha \sqrt{1-X_5^2}\widetilde{\mathrm{vol}}_3 \, ,\\
	&\psi_{\alpha+}= 
	 \hat y_\alpha \sqrt{1-X_5^2} 
	+ \hat y_\alpha \frac{e^{2A_5}}{9} X_5(1-X_5^2)\widetilde{\mathrm{vol}}_{\mathds{S}^2}+\pi \sqrt{1-X_5^2} \dd z \wedge \epsilon_{\alpha \beta \gamma} \hat y^\beta \dd \hat y^\gamma\,.
\end{split}	
\end{equation}
This is quite similar to (\ref{eq:psi7}), with a few changes: the factors of $4$ have become factors of $3$;
$X\to X_5 \equiv \frac{\alpha'}{\sqrt{\alpha'{}^2 - \frac32 \alpha \alpha''}}$; $A\to A_5$; as in (\ref{eq:ads5}), $\dd \psi\to D \psi$, with the tilde's denoting this change; and
\begin{equation}
	y^\alpha \to \hat y^\alpha\equiv y^\alpha|_{\varphi=\pi/2}=(0,\sin \theta, \cos \theta)
	\, ,\quad
	\dd y^\alpha \to \dd \hat y^\alpha = (-\sin \theta D \psi, \cos \theta \dd\theta , - \sin \theta \dd \theta)\,.
\end{equation}


\subsection{BPS probes} 
\label{sub:cal5}

The BPS condition for Mink$_4$ vacua is that the brane should be calibrated by $e^{-{\mathcal F}} \mathrm{Im} \Phi_-$, or equivalently that $e^{-{\mathcal F}} \mathrm{Re} \Phi_-|= 0$, $V\cdot e^{-{\mathcal F}} \Phi_+|=0$ $\forall \ V$ section of $T \oplus T^*$ \cite{Martucci:2005ht}. 

We are interested in brane probes that are extended along AdS$_5$ and are localized at a point in the internal space. The BPS condition becomes then
\begin{equation}\label{eq:cal5}
	e^{-{\mathcal F}} \psi_{0+}| =0\, ,\qquad e^{-{\mathcal F}} \psi_{\alpha+}|=0 \, \quad \alpha=1,2 
	 \,.
\end{equation}
This is formally identical to (\ref{eq:cal7}), although now it refers to the pure forms in (\ref{eq:psi5}).

The analysis of which branes are BPS is then formally identical to that of Sec.~\ref{sub:cal7}. The evaluation of the on-shell action is made more complicated by the presence of the external potentials (\ref{eq:Cext}). 

Just as discussed around (\ref{eq:alpha-max}), D4-branes are localized at the $z$ where $\alpha$ is maximal, and at $\theta=0$ or $\pi$. The DBI part of the action is very similar to (\ref{eq:SD4-7}):
\begin{equation}
	S_{\mathrm{D}4,\mathrm{DBI}}= T_4 \mathrm{Vol}({\mathcal M}_5) e^{5A-\phi}|_{\mathrm{D}4}= \mathrm{Vol}({\mathcal M}_5)\frac{\alpha_\mathrm{max}}{16\pi^4} \,.
\end{equation}
The WZ term now no longer vanishes:
\begin{equation}
	S_{\mathrm{D}4,\mathrm{WZ}}= T_4 \mathrm{Vol}({\mathcal M}_5) \frac13  \alpha_\mathrm{max} = \mathrm{Vol}({\mathcal M}_5) \frac{\alpha_\mathrm{max}}{48\pi^4}\,.
\end{equation}
Together these give
\begin{equation}\label{eq:SD4-5}
	S_{\mathrm{D}4}= \mathrm{Vol}({\mathcal M}_5) \frac{\alpha_\mathrm{max}}{12\pi^4} = \frac{27}{64} S_{\mathrm{D}4,\,\mathrm{AdS}_7}\,,
\end{equation}
where $S_{\mathrm{D}4,\,\mathrm{AdS}_7}$ is the result we obtained in (\ref{eq:SD4-7}) for the codimension-two brane probes in the AdS$_7$ solution. 

For the D6/D4 bound states, the DBI action and the quantized the charge become
\begin{equation}\label{eq:D6D4-DBI-q-5}
S_{\mathrm{D}6/\mathrm{D}4,\mathrm{DBI}}= \mathrm{Vol}({\mathcal M}_5)\frac1{32\pi^4} \int_{z_-}^{z_+} \alpha \dd z 
    \qquad m=\frac12(z_+-z_-)
\,.
\end{equation}
The D4 charge is the same as in \eqref{eq:qD6}. The evaluation of the WZ term from (\ref{eq:Cext}) is more laborious. First we compute the part without the world-volume flux:
\begin{align}
	\nonumber\int &(e^{-B} C)_7 = -\mathrm{Vol}({\mathcal M}_5)\frac23 \pi^2 \int (z \alpha - 2 \hat \alpha) \frac12 \dd \cos^2 \theta = \mathrm{Vol}({\mathcal M}_5)\frac23 \pi^2 \alpha_0^2\int \frac{\alpha'}{\alpha^3}(z \alpha - 2 \hat \alpha)\dd z \\
	&= \mathrm{Vol}({\mathcal M}_5) \frac23 \pi^2 \alpha_0^2\left[-\frac z \alpha + \frac{\hat \alpha}{\alpha^2}\right]_{z_-}^{z_+} 
	= \mathrm{Vol}({\mathcal M}_5)\frac23 \pi^2 \left(- \alpha_0 (z_+-z_-) + \int_{z_-}^{z_+} \alpha \dd z\right)\label{eq:eBC7}
\end{align}
In the second step we used (\ref{eq:acos}) to rewrite $\cos \theta$; in the fourth we recalled that $\alpha(z_-)= \alpha(z_+)= \alpha_0$. For the world-volume flux term, from (\ref{eq:qD6}) we get 
\begin{equation}
	-\int 2\pi f \wedge C_5= -2\pi \mathrm{Vol}({\mathcal M}_5) \int\frac13\cos \theta \alpha f_{z\varphi} \dd z = \mathrm{Vol}({\mathcal M}_5)\frac23 \pi^2 \alpha_0 (z_+-z_-)\,.
\end{equation}
The $z_+-z_-$ term cancels with that in $\int (e^{-{\mathcal F}}C)_7$. All in all, the on-shell brane action (\ref{eq:SDp}) reads
\begin{equation}\label{eq:SD6-5}
	S_{\mathrm{D}6/\mathrm{D}4} = \mathrm{Vol}({\mathcal M}_5) \frac1{24\pi^4}\int_{z_-}^{z_+} \alpha \dd z = \frac{27}{64} S_{\mathrm{D}6/\mathrm{D}4,\,\mathrm{AdS}_7} \,,
\end{equation}
where $S_{\mathrm{D}6/\mathrm{D}4,\,\mathrm{AdS}_7}$ is the result in (\ref{eq:SD6-7}). We see that the proportionality factor between the AdS$_7$ and AdS$_5$ results is $27/64$ both here and in (\ref{eq:SD4-5}); this will soon be important. 

For the massless case (\ref{eq:F0=0}), we now have
\begin{equation}\label{eq:SD-F0-5}
	S_{\mathrm{D}4}= \mathrm{Vol}({\mathcal M}_5)\frac{27}{64}\frac2{\pi^2}kN^2 \, ,\qquad
	S_{\mathrm{D}6/\mathrm{D}4}= \mathrm{Vol}({\mathcal M}_5)\frac{27}{64} \frac{2}{\pi^2}k m \left(N^2-\frac43 m^2\right)\,.
\end{equation}



\subsection{Probes of AdS solutions with Romans mass and D8-branes \label{sec:massivedefects}}

We now focus on some specific examples with non-zero Romans mass, also allowing for the presence of D8-branes and O8-planes. This will provide new interesting classes of defects and punctures in the dual field theories. As we have seen, we can consider BPS probes consisting of D4 and D4/D6 bound states in AdS$_7$ and AdS$_5$ solutions.
We will focus on the results \eqref{eq:SD4-5}, \eqref{eq:SD6-5} for the probes in AdS$_5$, but one could also use \eqref{eq:SD4-7}, \eqref{eq:SD6-7} for the AdS$_7$ probes. 

The first example that we consider is a class of solutions that is symmetric in the $z$ coordinate. It is regular at both poles $z=0$ and $z=N$, and it is divided into three regions $0 \leq z < \mu$, $\mu \leq z < N - \mu$, and $N-\mu \leq z \leq N$. Recall that $N$ corresponds to the $H$ flux quantum in the internal space. In these three regions the Romans mass takes value $2 \pi F_0=\{-n_0,0,n_0\}$ respectively. The jump in the Romans mass implies the presence of D8-brane stacks at $z=\mu$ and $z=N-\mu$. Their positions are dictated by the D6-charges they carry, which are $k$ and $-k$ respectively \cite{Apruzzi:2013yva}. The $F_2$ Bianchi identity fixes $\mu=\frac{k}{n_0}$. As for all solutions in this class, $-\alpha''/(81\pi^2)$ is piecewise linear, and its value at $z=i$ gives the gauge group ranks $r_i$ in the tensor branch of the dual SCFT:
\begin{equation}
   \begin{tikzpicture}[baseline, scale=1, thick, >=stealth]
  \def\muu{1.0}
  \def\NN{5.0}
  \def\HH{2.0} 

  \draw[->] (0,0) -- (\NN+0.5,0) node[right] {$z$};
  \draw[->] (0,0) -- (0,\HH+0.5) node[above] {$\alpha''(z)$};

  \draw[blue, thick, domain=0:\muu] plot(\x,{\HH/\muu*\x});
  \draw[blue, thick, domain=\muu:\NN-\muu] plot(\x,{\HH});
  \draw[blue, thick, domain=\NN-\muu:\NN] plot(\x,{\HH*(\NN-\x)/\muu});

  \draw[dashed] (\muu,0) -- (\muu,\HH) node[above left] {};
  \draw[dashed] (\NN-\muu,0) -- (\NN-\muu,\HH) node[above right] {};
  \node[below] at (0,0) {$0$};
  \node[below] at (\muu,0) {$\mu$};
  \node[below] at (\NN-\muu,0) {$N-\mu$};
  \node[below] at (\NN,0) {$N$};
\end{tikzpicture}
\end{equation}
\begin{equation} \label{eq:quivregD8}
\begin{tikzpicture}[font=\scriptsize, 
    baseline=(current bounding box.center),
    node distance=4mm, 
    roundnode/.style={circle, draw, thick, minimum size=9mm},
    squarednode/.style={rectangle, draw, thick, minimum size=7mm}
]
    \node[roundnode] (a1) {$n_0$};
    \node[roundnode, right=of a1] (a2) {$2n_0$};
    \node[right=of a2, xshift=0cm] (dots1) {$\dots$};
    \node[roundnode, right=of dots1, xshift=0cm] (ak1) {$k$};
    \node[roundnode, right=of ak1] (ak2) {$k$};
    \node[right=of ak2, xshift=0cm] (dots2) {$\dots$};
    \node[roundnode, right=of dots2, xshift=0cm] (ak3) {$k$};
    \node[roundnode, right=of ak3] (ak4) {$k$};
    \node[right=of ak4, xshift=0cm] (dots3) {$\dots$};
    \node[roundnode, right=of dots3, xshift=0cm] (a2_end) {$2n_0$};
    \node[roundnode, right=of a2_end] (a1_end) {$n_0$};

    \node[squarednode, below=of ak1] (b1) {$n_0$};
    \node[squarednode, below=of ak4] (b2) {$n_0$};

    \draw[thick] (a1) -- (a2);
    \draw[thick] (a2) -- (dots1);
    \draw[thick] (dots1) -- (ak1);
    \draw[thick] (ak1) -- (ak2);
    \draw[thick] (ak2) -- (dots2);
    \draw[thick] (dots2) -- (ak3);
    \draw[thick] (ak3) -- (ak4);
    \draw[thick] (ak4) -- (dots3);
    \draw[thick] (dots3) -- (a2_end);
    \draw[thick] (a2_end) -- (a1_end);

    \draw[thick] (ak1) -- (b1);
    \draw[thick] (ak4) -- (b2);
\end{tikzpicture}
\end{equation}
Regularity implies that $\alpha(z)$ is continuous and vanishes at the endpoints:
\begin{equation} \label{eq:regD8sol}
\alpha(z)=
\begin{cases}
 -\frac{27 \pi ^2 k}{2 \mu } z\left(z^2 + 3 \mu  (\mu -N)\right)
 &  0 \leq z < \mu\,;\\
\frac{27 \pi^2 k}{2} \left(3 z (N-z)-\mu^2\right) & \mu \leq z < N - \mu\,;\\
-\frac{27 \pi^2 k}{2 \mu } (N-z) \left( (N-z)^2+3 \mu(\mu-N)\right) & N-\mu \leq z \leq N \, .
\end{cases} 
\end{equation}
The on-shell action of the D4-brane probe in AdS$_5$ is given by \eqref{eq:SD4-5}; it reads explicitly
\begin{equation}
    S_{\rm D4}={\rm Vol}(\mathcal{M}_5) \frac{9 k }{32 \pi ^2}\left(3 N^2-4 \mu ^2\right)\,.
\end{equation}
The on-shell action of a D6/D4-probe contained in the massless region $\frac{N}{2}-m\leq z \leq \frac{N}{2}+m$, with $m<\frac{N}{2}-\mu$,
is given by \eqref{eq:SD6-5}, resulting in 
\begin{equation}
    S_{\rm D6/D4}={\rm Vol}(\mathcal{M}_5) \frac{8}{81\pi^4} \int_{\frac{N}{2}-m}^{\frac{N}{2}+m} \alpha(z) \dd z={\rm Vol}(\mathcal{M}_5) \frac{9km}{32 \pi ^2}
    \left(3 N^2-4 (\mu^2+m^2)\right)\, .
\end{equation}

Another interesting example has an O8-plane with a D8-brane source at $z=0$, and the sources for a stack of $M=\frac N{n_0}$ D6-branes at the pole $z=N$.  It has been discussed in \cite{Bah:2017wxp} (and generalized in \cite{Apruzzi:2017nck}). The SCFT was called there \emph{massive E-string}. The quiver in the tensor branch of this SCFT has linearly rising ranks:
\begin{equation} \label{eq:quivmassEstr}
\scalebox{0.9}{
\begin{tikzpicture}[font=\footnotesize,
    baseline=(current bounding box.center),
    node distance=8mm, 
    roundnode/.style={circle, draw, thick, minimum size=14mm},
    squarednode/.style={rectangle, draw, thick, minimum size=14mm}
]
    \node[squarednode] (e9n0) {$E_{9-n_0}$};
    \node[roundnode, right=of e9n0] (empty_node) {}; 
    \node[roundnode, right=of empty_node] (n0) {$n_0$};
    \node[roundnode, right=of n0] (2n0) {$2n_0$};
    \node[right=of 2n0] (dots) {$\dots$};
    \node[roundnode, right=of dots] (N_minus_1_n0) {\scriptsize $(N-1)n_0$};
    \node[squarednode, right=of N_minus_1_n0] (Nn0) {$N n_0$};

    \draw[thick] (e9n0) -- (empty_node);
    \draw[thick] (empty_node) -- (n0);
    \draw[thick] (n0) -- (2n0);
    \draw[thick] (2n0) -- (dots);
    \draw[thick] (dots) -- (N_minus_1_n0);
    \draw[thick] (N_minus_1_n0) -- (Nn0);

\end{tikzpicture}}    
\end{equation}
and correspondingly $-\alpha''/81\pi^2=n_0 z$, where $n_0=2\pi F_0$. The square node on the left denotes an enhanced flavor symmetry; the empty circular node denotes an E-string model.  We have
\begin{equation} \label{eq:massEstr}
    \alpha(z)= \frac{27}{2} \pi^2 n_0 \left(N^3-z^3\right)\,.
\end{equation}

The D4-brane probe has on-shell action in AdS$_5$ given by
\begin{equation}
    S_{\rm D4}={\rm Vol}(\mathcal{M}_5) \frac{27 n_0 N^3}{ 8 \pi ^2}
\end{equation}
whereas the on-shell action for the D6/D4 bound state probe extending from $z=0$ to $z=m$ reads
\begin{equation}
     S_{\rm D6/D4}={\rm Vol}(\mathcal{M}_5) \frac{8}{81\pi^4} \int_{0}^{m} \alpha(z) \dd z={\rm Vol}(\mathcal{M}_5) \frac{9 n_0 m}{64 \pi^2}\left(4N^3-m^3\right)\, .
\end{equation}

\subsection{Defect superconformal R-symmetry and anomaly} 
\label{sec:Rsym}
We now analyze the R-symmetry of codimension two defects in a 6d SCFT on $\mathbb{R}^6$ and on $\mathbb{R}^4\times$ a Riemann surface $\Sigma$.

Let us start by reviewing the 4d $\mathcal{N}=2$ R-symmetry $U(2)$. The superconformal R-symmetry charge of an $\mathcal{N}=1$ subalgebra $U(1)\subset U(2)$ has a generator \cite{Shapere:2008zf} 
\begin{equation} \label{eq:r5}
    r_5=\frac{4}{3} I_3 + \frac{1}{3}R_{\mathcal{N} =2}\,.
\end{equation}
$I_a$ with $a=1,2,3$ and $R_{\mathcal{N} =2}$ are the generators of the $\mathfrak{su}(2)$ and of the $\mathfrak{u}(1)$ algebra respectively. The name $r_5$ is chosen to remind us of its relevance for SCFT duals to AdS$_5$ solutions.

It can be verified that the combination \eqref{eq:r5} provides the correct R-symmetry assignment for free $\mathcal{N}=2$ multiplets \cite{Shapere:2008zf}. There is also a (weaker) group theoretical explanation for this particular combination.
We choose the $SU(2)\subset U(2)$ charges to be half-integer; the global structure of the R-symmetry group fixes the $U(1)$ charges to be double, i.e.~integers. So if an $\mathcal{N}=2$ multiplet has charge $1/2$ under $SU(2)$, it has charge $1$ under $U(1)$, and the combination \eqref{eq:r5} is a possible choice that gives integral charges under $r_5$. 

The superconformal anomalies $(a_5,c_5)$ can be expressed in terms of traces of these R-symmetry generators, which are the coefficient of the 4d anomaly polynomial,
\begin{equation}
    P_6 = \frac{k_{r_5r_5r_5}}{6}c_1(r_5)^3 - k_{r_5TT}\frac{c_1(r_5)p_1(TM_4)}{24}\,.
\end{equation}
$c_1(r_5)$ is the first Chern class of the superconformal R-symmetry bundle, $p_1(TM_4)$ is the first Pontryagin class of the 4d manifold where the 4d QFT and defects are defined, and
\begin{equation}
    k_{r_5r_5r_5}= {\rm tr}(r_5^3), \qquad k_{r_5TT}= {\rm tr}(r_5)\,.
\end{equation}
The superconformal anomalies are written in terms of the coefficients \cite{Anselmi:1997am}
\begin{equation}
    a_5= \frac{3}{32}(3{\rm tr}(r_5^3) -{\rm tr}(r_5)), \qquad c_5= \frac{1}{32}(9{\rm tr}(r_5^3) -5{\rm tr}(r_5))\,.
\end{equation}
In addition, any $\mathcal N=2$ SCFT satisfies \cite{Shapere:2008zf,Kuzenko:1999pi}
\begin{equation}
   {\rm tr}(R_{\mathcal N=2}^3)= {\rm tr}(R_{\mathcal N=2})= 48 (a_5-c_5), \quad  {\rm tr}(R_{\mathcal N=2}I_aI_b)=  \delta_{ab}(4a_5 - 2c_5). 
\end{equation}
For holographic SCFTs, $a_5=c_5$ at leading order in the large $N$ expansion, which means that $ {\rm tr}(R_{\mathcal N=2}^3)= {\rm tr}(R_{\mathcal N=2})={\rm tr}(r_5)=0$; in particular,
\begin{equation}
    {\rm tr}(r_5^3) = \frac{16}{9} {\rm tr}(I_3^2 R_{\mathcal{N}=2}) = \frac{32}{9} a_5\,.
\end{equation}

When we obtain an $\mathcal{N}=2$ 4d theory as a compactification of a 6d SCFT on $\Sigma$, we identify
\begin{equation} \label{eq:64dRsymid}
    I_3^{(6d)}=I_3, \qquad K_{U(1)_{\Sigma}} = R_{\mathcal N =2 }\,.
\end{equation}
$I_a^{(6d)}$ are the $\mathcal{N}=(1,0)$ R-symmetry generators, and the $U(1)_{\Sigma}$ is the structure group of the canonical bundle of $\Sigma$ identified with $R_{\mathcal N=2}$. This is also valid in the presence of punctures on $\Sigma$ that preserve $\mathcal N=2$ in 4d. As we saw in the introduction, we will also view them as codimension-two defects of the 6d $\mathcal{N}=(2,0)$ SCFT on $\Sigma$, that are pointlike on $\Sigma$.

It is also possible to compactify 6d $\mathcal{N}=(2,0)$ SCFT on a $\Sigma$ in a way that only preserves $\mathcal N=1$ in 4d \cite{Maldacena:2000mw,Bah:2012dg}. (As mentioned in Sec.~\ref{sub:ads5}, the $F_0=0$ case of the AdS$_5$ solutions in this section are actually of this type.) The superconformal R-symmetry is given by
\begin{equation} \label{eq:r5p}
        r_5'= I_3 + \frac{1}{2}R_{\mathcal{N} =2}\,.
\end{equation}
The $\mathcal{N}=1$ 4d SCFT is related to the $\mathcal{N}=2$ 4d SCFT by an RG flow \cite{Tachikawa:2009tt} which for Lagrangian theories is triggered by adjoint mass deformation. $r_5'$ in \eqref{eq:r5p} is the combination that is preserved by the adjoint mass deformation. For holographic theories this leads to the following property between the anomaly polynomial coefficients and the conformal anomaly $a'_5$:
\begin{equation}
    {\rm tr}(r_5'^3) =\frac{3}{2} {\rm tr}(R_{\mathcal N =2} I_3^2) = \frac{32}{9} a'_5,
\end{equation}
which leads to the following famous relation \cite{Tachikawa:2009tt}
\begin{equation}\label{eq:27/32}
     a'_5 = \frac{27}{32} a_5.
\end{equation}
Similarly to the $\mathcal N =2$ case, we can also include punctures in the story. 

So far we have discussed anomalies of a 6d SCFT compactified on $\Sigma$,
where we identified the 6d R-symmetry subalgebra and the canonical bundle of the Riemann surfaces with the 4d $U(2)$ R-symmetry generators of the 4d SCFT. We now would like to go back and see the relation of the conformal anomalies $a_5$ and $a_5'$ with the one for a defect of a 6d SCFT with no Riemann surface. In this case we cannot rely on any identification of the 6d R-symmetry generators with the ones for the $U(2)$ R-symmetry of the 4d SCFT. We still want to look at a codimension-two defect that preserves $\mathcal{N}=1$ in 4d and compute the defect conformal anomaly $a_7$. The symmetry of the system is now $SO(3) \times U(1)$. The $SO(3)$ directly descends from the 6d theory---we generically consider $\mathcal{N}=(1,0)$. Since $SO(3)$ cannot be embedded in $U(2)$, the $SO(3)$ charges will be normalized like the $U(1)$ charges. The $U(1)$ comes from the space orthogonal to the defect, whose generator we call $K_{U(1)}$. If we normalize the charges to be half-integral, a combination, also proposed in \cite[(4.33)]{Wang:2021mdq}, that gives integral charges for the superconformal R-symmetry is now given by
\begin{equation}
    r_7= \frac{2}{3}(2 I_3^{(6d)}+ K_{U(1)})\, .
\end{equation}
 If we now formally rely on the identification \eqref{eq:64dRsymid}, we get the following useful relation, 
\begin{equation}\label{eq:64/27}
    a_7 =2 a_5= \frac{64}{27} a_5'
\end{equation}
for a superconformal codimension-two defect of a 6d SCFT preserving $\mathcal{N}=1$ superconformal algebra.

In the next section, we will relate the $a$ anomaly for defects to the on-shell brane action. Thanks to this fact, the factor $64/27$ in \eqref{eq:64/27} can be viewed as an explanation for the $27/64$ in \eqref{eq:SD4-5}, \eqref{eq:SD6-5}.

\section{Anomaly coefficients} 
\label{sec:an}

As mentioned in the introduction, the contribution of a codimension-two defect to the Weyl anomaly of a 6d SCFT can be parameterized as $\delta(D) \frac1{(4\pi)^2} (-a_{D} E_4 + \sum d_I J_I)$ \cite[(3.1)]{Chalabi:2021jud}. We will now show how to compute holographically the coefficient $a_D$ (which we just call $a$ throughout this section) and $d_2$. As mentioned above, all of the (parity-even) defect Weyl anomalies are related, expressible in terms of the brane tension.  So once we have computed $a$ from the on-shell action of a probe brane dual to a spherical defect, the computation of $d_2$ using the holographic entanglement entropy (EE) of a spherical region will provide verification of the ratio between $a$ and the other defect Weyl anomalies. This comes in addition to the physical information about EE and the stress-energy tensor one-point function that it carries.

\subsection{\texorpdfstring{Sphere partition function and $a$ coefficient}{Sphere partition function and a-coefficient}} 
\label{sub:a}

In this subsection, we will use probe brane holography to compute the A-type defect Weyl anomaly coefficient $a$ for the codimension-two defects in Sec.~\ref{sec:ads7} and \ref{sec:ads5}.  From the solutions holographically dual to the probe limit of flat co-dimension two defects in $\mathbb{R}^{5,1}$ (or $\mathbb{R}^6$), we can then use the broken conformal generators at the boundary $\partial\CM_5\hookrightarrow \partial{\rm AdS}_7$ to map the boundary geometry to $\mathds{S}^4\hookrightarrow \mathbb{R}^6$.  These transformations can then be extended into the bulk where $\mathcal{M}_5\to \widetilde{\mathcal{M}}_5$ is embedded in  AdS$_7$ with the appropriate boundary geometry. This computation is a straightforward extension of the conformal transformation in \cite[App. A]{Apruzzi:2024ark}, which we will briefly recapitulate for completeness here.  

We begin by writing the flat boundary geometry $\mathbb{R}^6$ with Euclidean metric
\begin{align}
{\rm d}s^2 = \delta_{\mu\nu}{\rm d}x^\mu{\rm d}x^\nu,
\end{align}
where $\mu,\nu = 1,\ldots, 6$. After using the broken conformal generators to map from a flat to spherical defect, the defect submanifold is defined by $\{x_1^2+\ldots x_5^2=L^2, x^6=0\}$, which is conveniently described by the map to polar coordinates:
\begin{align}
x^1 = \varrho \cos\zeta,\quad x^2 = \varrho \sin\zeta\cos \chi_1,\quad\ldots\quad\,x^5 = \varrho \sin\zeta \sin\chi_1\sin\chi_2\sin\chi_3\,.
\end{align}
We end up with a boundary geometry described by the line element
\begin{align}\label{eq:polar-metric}
{\rm d}s^2 = {\rm d}\varrho^2 + \varrho^2{\rm d}\Omega_4^2 + \dd y^2,
\end{align}
relabeling $x^6= y$. Then, by the coordinate transformation
\begin{align}\label{eq:H5-S1-transformation}
\varrho = \frac{ L \sinh\rho}{\cosh\rho - \cos\varphi},\qquad y = \frac{L\sin\varphi}{\cosh\rho-\cos\varphi},
\end{align}
with $\varphi\sim\varphi+2\pi$, we end up, up to an overall conformal factor, with flat space described by the line element
\begin{align}
{\rm d}s^2 = L^2({\rm d}\rho^2 + {\rm d}\varphi^2+ \sinh^2\rho~{\rm d}\Omega_4^2).
\end{align}
The boundary geometry is now $\mathds{H}^5\times \mathds{S}^1$.  The defect wraps the $\mathds{S}^4$ at the boundary of $\mathds{H}^5$, $\rho\to\infty$. 

At the level of the on-shell action, the net effect is to change the structure of the UV divergences to those associated with the volume of $\widetilde{\mathcal{M}}_5$.  Recall that the on-shell action took the form
\begin{equation}
	S_{\rm brane}= \mathrm{Vol}({\widetilde{\mathcal M}}_5) {\mathcal I}\,;
\end{equation}
for AdS$_5$ probes. In \eqref{eq:SD4-5}, \eqref{eq:SD6-5} we obtained $\mathcal{I}= \alpha_{\rm max}/12\pi^2$ for the D4 branes and  $24\pi^2\mathcal{I} = \int_{z_-}^{z_+} \alpha {\rm d} z$ for the D6/D4 bound state in the AdS$_5\times \Sigma\times M_3$ solutions.  Following the transformation that yields $\partial\widetilde{\mathcal{M}}_5=\mathds{S}^4$, and imposing a UV cutoff at $\rho =\log R\Lambda$ the bare volume of $\mathds{H}^5$ is $\mathrm{Vol}(S^4)\int_0^{\log R\Lambda}\sinh^4\rho\dd \rho \sim \frac{8\pi^2}3 (\frac1{64}(R\Lambda)^4 - \frac18 (R\Lambda)^4+ \frac38\log(R\Lambda)+\ldots)$. The polynomial divergences
can be removed by standard counterterms in holographic renormalization \cite{Karch:2005ms}. The remaining, physical, log-divergent part of the on-shell action is then
\begin{align}
\mathrm{Vol}(\widetilde{\mathcal{M}}_5) = \pi^2\log R\Lambda +\ldots \qquad \Rightarrow \qquad S_{\rm on-shell}= \pi^2 \mathcal{I}~\log R\Lambda + \ldots~.
\end{align}
We have suppressed the subleading, scheme-dependent finite terms. 
Recall that the anomalous constant Weyl rescaling transformation of the sphere free energy is computed using
\begin{align}
\partial_{\log R\Lambda} \log Z_{\mathds{S}^4} = - \int_{\mathds{S}^4}\sqrt{g_{\mathds{S}^4}}\left<T^{\mu}{}_\mu\right>~.
\end{align}
The trace anomaly for the flat embedding $\mathds{S}^4\hookrightarrow\mathbb{R}^6$, i.e.~with vanishing second fundamental form, only receives contributions from the A-type anomaly. Hence, the remaining integral is found to be
\begin{align}
\partial_{\log R\Lambda} \log Z_{\mathds{S}^4}=\frac{1}{16\pi^2}a\int_{\mathds{S}^4}\sqrt{g_{\mathds{S}^4}}E_4 = 4a~,
\end{align}
where we have used the normalization of the defect trace anomaly in \cite[(3.1)]{Chalabi:2021jud} and $\int_{\mathds{S}^4}E_4 = 64\pi^2$. Thus, we find that the holographic A-type defect Weyl anomaly coefficients are given by
\begin{equation}\label{eq:a-brane}
	a= \frac{\pi^2}4 {\mathcal I}\,.
\end{equation}
For example, applying this to the D4 and D6/D4 on-shell actions in the AdS$_5$ backgrounds (\ref{eq:SD4-5}), (\ref{eq:SD6-5}) gives
\begin{equation}\label{eq:aD4D6}
	a_{\mathrm{D}4}= \frac{\alpha_\mathrm{max}}{48\pi^2} \, ,\qquad
	a_{\mathrm{D}6/\mathrm{D}4} = \frac1{96\pi^2}\int_{z_-}^{z_+} \alpha \dd z\,.
\end{equation}
In particular, for the $F_0=0$ solution (\ref{eq:F0=0}) we obtain
\begin{equation}\label{eq:aD4D6-0}
	a_\mathrm{D4}= \frac{27}{32}k\frac{N^2}4 \, ,\qquad a_{\mathrm{D}6/\mathrm{D4}}= \frac{27}{32}k\frac m4 \left(N^2-\frac43 m^2\right)\,.
\end{equation}

As a brief consistency check, consider the Graham--Reichert energy for an AdS$_5$ probe brane \cite{Graham:2017bew}.  The A-type anomaly is determined to be $a = \pi^2 L^5 T_{\rm brane} / 4$ \cite{Chalabi:2021jud}, which is precisely the form in \eqref{eq:a-brane} with $\mathcal{I} \equiv L^5 T_{\rm brane}$.  The subtlety here is in the definition of the effective AdS length $L^5$.  If we write $L^5 = e^{5A-\phi}$ and integrate over the internal  $\{z,\theta\}$ directions, we find an exact match from the Graham--Reichert energy to the expressions above.  Thus, we also have a prediction for all 23 parity-even probe brane Weyl anomaly coefficients since they are all proportional to $\pi^2 L^5 T_{\rm brane} = \pi^2\mathcal{I}$, see the co-dimension $q=2$ values in \cite[Table 1]{Chalabi:2021jud}.  


\subsection{Entanglement entropy} 
\label{sub:ee}
In this subsection, we are interested in computing the defect contribution to the holographic entanglement entropy (EE) of a spherical region in the boundary theory.  In particular, our goal is to compute the universal, log divergent part of the sphere EE.  The universal part of the defect sphere EE was shown in \cite[(4.38)]{Chalabi:2021jud} to be a linear combination of two defect Weyl anomalies:
\begin{align}\label{eq:defect-sphere-ee}
S_{\text{EE, defect}}|_{\log} = - \frac{2}{5}\left(10a + d_2 \right)\log\frac{R}{\epsilon}~.
\end{align}
In a generic $d\geq6$ dimensional CFT with a four-dimensional defect, it was also shown in \cite{Chalabi:2021jud} that the B-type defect Weyl anomaly coefficient $d_2$ fixes the one-point function of an insertion of the stress-energy tensor in the ambient space away from the defect.   The appearance of $d_2$ in $S_{\rm EE, defect}$ thus originates from the Killing energy for the time translation Killing vector, while the appearance of $a$ originates from the sphere free energy.  Combining with results in the previous subsection for the A-type anomaly, we can then compute $d_2$ from the defect sphere EE.

To compute the sphere EE, we follow the map to the hyperbolic black hole  by writing the AdS$_7$ geometry (setting the AdS length $L=1$) as
\begin{subequations}
\begin{align}
{\rm d}s_{\rm AdS_7}^2 &= {\rm d}x^2 + \sinh^2x~{\rm d}\varphi^2 + \cosh^2x~{\rm d}s_{\mathcal{M}_5}^2~,\\
{\rm d}s_{\mathcal{M}_5}^2&= \frac{1}{Z^2}({\rm d}Z^2 - {\rm d}t^2+{\rm d}r_\parallel^2 + r_\parallel^2 {\rm d}\Omega_2^2)~,
\end{align}
\end{subequations}
where the probe brane wraps $\{Z, t, r_\parallel, {\mathds{S}}^2\}$. Then by mapping \cite{Casini:2011kv,Jensen:2013lxa}
\begin{align}\begin{split}
Z&= \frac{R}{\nu \cosh\rho + \cosh\tau \sqrt{\nu^2-1}}~,\\
 r_\parallel &= \frac{R\nu\sinh\rho}{\nu \cosh\rho + \cosh\tau \sqrt{\nu^2-1}}~,\\
 t &= \frac{R\sinh\tau \sqrt{\nu^2-1}}{\nu \cosh\rho + \cosh\tau \sqrt{\nu^2-1}},~
\end{split}\end{align}
the geometry on $\mathcal{M}_5$ becomes the hyperbolic black hole
\begin{align}
{\rm d}s_{\mathcal{M}_5}^2 = \frac{{\rm d}\nu^2}{f(\nu)} - f(\nu){\rm d}\tau^2 + \nu^2({\rm d}\rho^2 +\sinh^2\rho~{\rm d}\Omega_2^2)
\end{align}
with $f(\nu) = \nu^2-1$ and periodic boundary conditions around  $\tau \sim \tau+2\pi$. This can be generalized to the finite temperature $\beta = 2\pi n$ background
\begin{align}
f(\nu) = \nu^2 -1 + \frac{\nu_h^4}{\nu^4}(\nu_h^2-1),\qquad \nu_h \equiv \frac{1}{3\beta}(\pi +\sqrt{\pi^2+6\beta^2})~;
\end{align}
now the periodicity condition reads $\tau  \sim \tau + \beta$.  

Embedding D4 probe branes or D6/D4 bound states in this background, the probe action reads
\begin{align}
S_{\rm brane} = \frac{\pi}{4}\beta(\nu_c^4 - \nu_h^4)(\sinh(2\rho_c)-2\rho_c) \mathcal{I}~,
\end{align}
where we have introduced two radial cutoffs: $\rho_c$ along $\mathds{H}^3$ and $\nu_c$ on AdS$_5$. Holographically renormalizing away the divergences in the AdS$_5$ radial direction, writing $\rho_c = \log\frac{2R}{\epsilon}$, and expanding around the entangling surface $\epsilon\to 0$, the renormalized probe brane action takes the form
\begin{align}
S_{\rm ren} = \frac{\pi}{16}\beta(8\nu_h^4-3)\mathcal{I}\left(-\frac{R^2}{\epsilon^2} + \log\frac{2R}{\epsilon}+\ldots\right)~.
\end{align}
The leading quadratic divergence is a non-universal shape-dependent contribution. Computing the R\'enyi entropy and taking the $n\to 1$ limit, this renormalized probe brane action maps on to the defect sphere free energy by $S_{\rm EE}= \beta^2\partial_\beta (\beta^{-1}S_{\rm ren})|_{\beta\to2\pi}$ \cite{Hung:2011nu}. This yields the universal coefficient of the log divergent part
\begin{align}
S_{\rm EE}|_{\rm univ} = -\frac{4}{5}\pi^2\mathcal{I}~.
\end{align}
Then using the value of $a =\pi^2\mathcal{I}/4$ and (\ref{eq:defect-sphere-ee}), we find
\begin{align}
d_2 = -\frac{\pi^2}{2}\mathcal{I}~.
\end{align}
As a quick check, comparing to \cite[Table 1]{Chalabi:2021jud}, we find exact agreement with the Graham--Reichert energy computation; for the AdS$_5$ probe brane embedded in AdS$_7$, the relative factor between $a$ and $d_2$ is $-2$.  So, for the D4 and D6/D4 solutions in (\ref{eq:SD4-5}), (\ref{eq:SD6-5}) we find
\begin{align}
d_2{}^{\rm D4} = -\frac{\alpha_{\rm max}}{24},\qquad d_2{}^{\rm D6/D4} = -\frac{1}{48\pi^2}\int_{z_-}^{z_+}\alpha {\rm d} z~,
\end{align}
which for the $F_0=0$ solutions (\ref{eq:F0=0}) read
\begin{align}
d_2{}^{\rm D4} = -\frac{27}{32}k\frac{N^2}{2},\qquad d_2{}^{\rm D6/D4} = -\frac{27}{32}k\frac{m}{2}\left(N^2 - \frac{4}{3}m^2\right)~.
\end{align}

Aside from the defect sphere EE, we can compute the defect R\'enyi entropies as well. We skip the details, as they follow the same logic but on the $n$-fold branched cover ($\beta \sim \beta + 2\pi n$); this gives the universal coefficient of the $n^{\rm th}$ defect Renyi entropy as
\begin{align}
S^{(n)}_{\rm univ} = \frac{\pi^2\mathcal{I}}{162n^3(n-1)}\left(1-90n^4+24n^2 +(1+12n^2)\sqrt{1+24n^2}\right)~.
\end{align}
One can see that the $n\to 1$ limit recovers the defect EE. An interesting limits to consider is  $n\to0$, whose leading coefficient (modulo sign) counts the number of non-zero eigenvalues of the reduced density matrix
\begin{align}
\lim_{n\to 0}S^{(n)}_{\rm univ} = -\frac{\pi^2\mathcal{I}}{81n^3} + O(n^{-2})~,
\end{align}
which in the specific D4 and D6/D4 solutions give
\begin{align}
S^{(0)}_\mathrm{D4} = - \frac{\alpha_{\rm max}}{972n^3}+ O(n^{-2})\,,\qquad S^{(0)}_\mathrm{D6/D4} =- \frac{1}{1944\pi^2n^3}\int_{z_-}^{z^+}\alpha {\rm d}z~,
\end{align}
and  in the $F_0=0$ case
\begin{align}
S^{(0)}_{\mathrm D4} = - \frac{27}{32}\frac{kN^2}{81n^3}+ O(n^{-2})\,,\qquad S^{(0)}_\mathrm{D6/D4} =-\frac{27}{32} \frac{km}{81n^3}\left(N^2 - \frac{4}{3}m^2\right),
\end{align}
Finally, the limit $n\to\infty$ measures the largest eigenvalues of the reduced density matrix
\begin{align}
\lim_{n\to\infty} S^{(n)} = - \frac{5}{9}\pi^2\mathcal{I} + O(n^{-1}).
\end{align}
The values for the specific cases of the D4 and D6/D4 solutions are obtained with the same substitutions as above.

\subsection{\texorpdfstring{$\mathds{S}^1\times\mathds{S}^3$ partition function}{S1xS3 partition function}}

In this subsection, we consider the transformation where the defect wraps $\mathds{S}^1_\beta\times\mathds{S}^3\hookrightarrow \mathds{S}^1_\beta\times\mathds{S}^5$ in the boundary field theory.  The partition function of the boundary theory on the $\mathds{S}^1_\beta\times\mathds{S}^5$ is related to the superconformal index $I$ by
\begin{align}
\mathcal{Z}_{\mathds{S}^1\times \mathds{S}^5} = e^{-\beta \mathcal{E}_\mathrm{C}}I~.
\end{align}
The term $\mathcal{E}_\mathrm{C}$ in the prefactor is the \emph{supersymmetric Casimir energy} (SCE), which is expressible in terms of the anomalies of the theory \cite{Assel:2015nca, Bobev:2015kza, Martelli:2015kuk}.  Changing the ambient theory by adding defect degrees of freedom modifies the structure of the anomalies, and hence $\mathcal{E}_\mathrm{C}$. This has been demonstrated for two-dimensional boundaries \cite{Bullimore:2020jdq} and defects \cite{Chalabi:2020iie, Huang:2025eze}. The $O(1)$ part of the holographically renormalized on-shell probe brane action thus computes the change in $\mathcal{E}_\mathrm{C}$ due to the defect degrees of freedom in the probe limit. Though a precise form for the defect SCE in terms of anomalies for four-dimensional defects is not yet known, the goal of this computation is to provide a holographic prediction, which can provide a clue to how the defect SCE can be expressed in terms of defect anomalies.

The map from flat space to $\mathds{S}^1_\beta\times \mathds{S}^5$ starts by the same map to (\ref{eq:polar-metric}). As before we take the defect to lie at $\{\varrho^2+y^2 = L^2\}$, but now instead of mapping to $\mathds{S}^1\times\mathds{H}^5$ using (\ref{eq:H5-S1-transformation}), we use
\begin{align}\label{eq:s1s3-trafo}
\varrho = \frac{L\sin\varsigma}{\cosh\tau -\cos\varsigma},\qquad y = \frac{L\sinh\tau}{\cosh\tau - \cos\varsigma}~,
\end{align}
which up to a conformal factor brings us to $\mathbb{R}\times \mathds{S}^5$;
\begin{align}
{\rm d}s^2 = L^2({\rm d}\tau^2+{\rm d}\varsigma^2 + \sin^2\varsigma~{\rm d}\Omega_4^2).
\end{align}
Compactifying $\tau\sim\tau+\beta$ yields a $\mathds{S}^1_\beta\times\mathds{S}^5$, which can be extended into the bulk
\begin{align}
{\rm d}s_7^2 = \dd\sigma^2 + \cosh^2\sigma {\rm d}\tau^2 + \sinh^2\sigma {\rm d}\Omega_5^2~.
\end{align}
Under the transformation in (\ref{eq:s1s3-trafo}), the defect is now mapped to the surface at $\zeta = \varsigma = \frac{\pi}{2}$.  Extending into the bulk, we see that the probe brane wraps the holographic radial coordinate $\sigma$, $\mathds{S}^1_\beta$, and an $\mathds{S}^3\hookrightarrow\mathds{S}^5$ on the equator.

We again forego the details of the calculation, as it follows the same process of computing the probe brane action and holographically renormalizing, as above. We find that the leading constant part of the action takes the form
\begin{align}
S_{\rm ren} = \frac{3}{32}\beta\pi^2\mathcal{I}~.
\end{align}
So, the defect SCE is
\begin{align}
\mathcal{E}_\mathrm{C}^{\rm (def)} =-\frac{3}{32}\pi^2\mathcal{I}~.
\end{align}
Evaluating on the D4 and D6/D4 solutions we find
\begin{align}
\mathcal{E}^{\rm (def)}_\mathrm{C,D4} =- \frac{ \alpha_{\rm max}}{128},\qquad \mathcal{E}^{\rm (def)}_\mathrm{C,D6/D4}  = -\frac1{256}\int_{z_-}^{z_+} \alpha {\rm d} z~.
\end{align}
In the $F_0=0$ limit, we get
\begin{align}
\mathcal{E}^{\rm (def)}_\mathrm{C,D4} = -\frac{27}{32}\frac{3}{32}k N^2,\qquad \mathcal{E}^{\rm (def)}_\mathrm{C,D6/D4} = - \frac{27}{32}\frac{3}{32}km\left(N^2 - \frac{4}{3}m^2\right).
\end{align}



\section{Small Punctures in Class S}
\label{sec:s}

Our main task in this section is to verify that the $a$ anomaly coefficients for our AdS$_5$ defects match the contributions from punctures in the particular case of class S theories. We discuss the situation for the orbifold class S$_k$ in Sec.~\ref{sub:Sk}. Finally in Sec.~\ref{sub:beyond} we spell out our gravity predictions for punctures for two particular $\mathcal{N}=(1,0)$ theories.  

\subsection{Theories of Class S}
\label{sub:s}
Theories of class S are 4d $\mathcal{N}=2$ SCFTs realized by compactifying the 6d $\mathcal{N}=(2,0)$ theory of type $A_{N-1}$ on a Riemann surface $\Sigma$ with regular punctures \cite{Gaiotto:2009we, Gaiotto:2009hg}. Each puncture corresponds to a half-BPS codimension-two defect of the 6d theory, extended over the 4d spacetime and localized at a point on $\Sigma$; it is characterized by a partition of $N$. We denote a partition by
\bes{
\rho = [w_p^{k_p}, w_{p-1}^{k_{p-1}}, \ldots, w_2^{k_2}, w_1^{k_1}]~,
}
such that
\bes{ \label{eq:partN}
N = \sum_{a=1}^p w_a k_a~, \quad 1\leq w_1 < w_2 < \cdots < w_p~, \quad k_i \geq 1~\text{ for all $i=1, \ldots, p$}~.
}
This can be represented by a Young diagram as depicted in \cite[Fig.~3]{Bah:2019jts}. In the holographic dual \cite{Gaiotto:2009gz}, the puncture consists of $p-1$ M5-brane stacks, the $a$-th ($a=1,\ldots,p-1$) consisting of $w_a$ M5s; if $k_a>1$, that point is a $\mathds{Z}_{k_a}$ singularity.\footnote{This identification can be read off from \cite[Sec.~3.4]{Gaiotto:2009gz}. In the notation of that paper, the number of M5-branes is related there to the differences $\lambda_I-\lambda_{I-1}$ in the intercepts of the charge densities $\lambda=\lambda_I + r_I \eta$; these differences can then be seen to be equal to the $w_a$ above, in our language.} Upon reduction to IIA, one expects these to become bound states of $k_a$ D6-branes and $w_a$ D4-branes.

It is also useful to define the transpose of the partition, $\rho^t$, as
\bes{ \label{transposepart}
\rho^t = [\ell_1^{w_1}, \ell_2^{w_2-w_1}, \ell_3^{w_3-w_2}, \ldots, \ell_p^{w_p-w_{p-1}}]~,
}
where
\bes{
N = \sum_{a=1}^p (w_a-w_{a-1}) \ell_a~, \quad 1\leq \ell_1 < \ell_2 < \cdots < \ell_p~.
}
Following \cite[Sec.~6.2]{Bah:2019jts}, we rewrite the partition $\rho^t$ as $(\tilde{\ell}_1, \tilde{\ell}_2, \ldots, \tilde{\ell}_{\tilde{p}})$,\footnote{Here we use the round brackets to emphasize the difference between this way of writing the partition and \eqref{transposepart}.} where repetitions are allowed, such that
\bes{
N = \sum_{i=1}^{\tilde{p}} \tilde{\ell}_i~, \quad \tilde{p} = w_p~, \quad 1\leq \tilde{\ell}_1 \leq \tilde{\ell}_2 \leq \cdots \leq \tilde{\ell}_p~.
}
The Young diagram for $\rho^t$ is shown in \cite[Fig.~4]{Bah:2019jts}. We use the indices $a=1, \ldots, p$ and $i = 1, \ldots, \tilde{p}$, and define the additional quantities
\bes{ \label{eq:furtherdef}
\begin{array}{ll}
\ell_a =\sum_{b=a}^p k_b~, &\qquad N_a =\sum_{b=1}^a (w_b - w_{b-1})\ell_b = w_a \ell_a +\sum_{b=1}^{a-1} w_b k_b \\
\tilde{N}_i = \sum_{j=1}^i \tilde{\ell}_j~, &\qquad \tilde{k}_i = \tilde{\ell}_i - \tilde{\ell}_{i+1}~~ \text{with $\tilde{\ell}_{\tilde{p}+1} =0$}~.
\end{array}
}
As discussed in \cite[(6.27)]{Bah:2019jts} and above \cite[(6.21)]{Bah:2022yjf}, these definitions are related by
\bes{ \label{eq:tildeandnontilde}
\tilde{k}_i=k_a~, \quad \tilde{N}_i =N_a~, \quad \text{if $i=w_a$}.
}

\subsection{Anomalies}
\label{sub:a-s}

The $a$ anomaly of these theories is a sum of contributions from the bulk and the defects:
\begin{equation}\label{eq:a-tot}
	a = \frac{1}{48}{(2g-2)}(N-1) (8 N^2 +8 N +5) + \sum_\rho a_\rho\,,
\end{equation}
where the contribution from a puncture $\rho$ is given by
\begin{equation} \label{eq:arho1}
	a_\rho = \frac{1}{48}N(8N^2-6N-3)-\frac{1}{4} \sum_i (N^2- \tilde N_i^2) +\frac{1}{48}\sum_a N_a k_a\,.
\end{equation}
This formula can be derived from \cite[(6.16), (6.17)]{Bah:2019jts} as follows. First of all, we remark that our notation differs slightly from that in \cite{Bah:2019jts}, as our bulk contribution is proportional to $(2g-2)$ rather than $(2g-2+n)$, where $n$ is the number of punctures. Denoting the defect contribution in \cite{Bah:2019jts} as $\hat a_\rho$,\footnote{For a completely closed (trivial) puncture $\rho=[N]$, we have $a_\rho=0$ and $\hat a_\rho=\frac{1}{48}(N-1) (8 N^2 +8 N +5)$.} the relationship is
\bes{
a_\rho &= \frac{1}{48}(N-1) (8 N^2 +8 N +5)+ \hat a_\rho \\
&= \frac{1}{48}(N-1) (8 N^2 +8 N +5) + \frac{1}{24}( n_h(\rho)+5 n_v(\rho))\\
&= \frac{1}{48}(N-1) (8 N^2 +8 N +5) + \frac{1}{24}\left[6n_v(\rho) -(n_v(\rho)-n_h(\rho)) \right]~,
}
where $n_v(\rho)$ and $n_h(\rho)$ are the effective numbers of vector multiplets and hypermultiplets, respectively, written in the convention of \cite{Bah:2019jts}. Substituting the expressions for $n_v(\rho)-n_h(\rho)$ and $n_v(\rho)$ from \cite[(6.16), (6.17)]{Bah:2019jts} yields \eqref{eq:arho1} as required.\footnote{We replaced the sum $\sum_i \tilde{N}_i \tilde{k}_i$ with $\sum_a N_a k_a$ via \eqref{eq:tildeandnontilde}. We also used the identity $\frac{1}{48}N(8N^2-6N-3) = \frac{1}{48}(N-1) (8 N^2 +8 N +5) + \frac{1}{48}(5-6N^2)$.} 

Let us now proceed further. Using the identity from \cite[(6.21)]{Bah:2022yjf},\footnote{We note that the term $\frac{1}{2}$ in \cite[(6.21)]{Bah:2022yjf} should not be present.}
\bes{
\scalebox{0.9}{$
\begin{split}
\sum_i (N^2- \tilde N_i^2) &= \sum_{a=1}^{p} \left( \frac{2\ell_a^2}{3}(w_a^3 - w_{a-1}^3) + \ell_a(N_a - w_a\ell_a)(w_a^2 - w_{a-1}^2) - \frac{N_a k_a}{6} \right) -\frac{N^2}{2}~,
\end{split}$}
}
we obtain another expression for the anomaly (see also \cite[(6.2)]{Bah:2019jts}):
\bes{
\scalebox{0.87}{$
\begin{split}
a_\rho = \frac{1}{48}N(8N^2-3) -\frac{1}{4} \sum_{a=1}^{p} \left( \frac{2}{3} \ell_a^2 (w_a^3 - w_{a-1}^3) + \ell_a (N_a - w_a \ell_a) (w_a^2 - w_{a-1}^2) - \frac{1}{4} N_a k_a \right)~.
\end{split}$}
}
Using the relations from \eqref{eq:partN} and \eqref{eq:furtherdef},
\bes{
\begin{array}{ll}
\ell_a =\sum_{b=a}^p k_b~, &\qquad N_a = w_a \ell_a +\sum_{b=1}^{a-1} w_b k_b~,\\
N=\sum_{a=1}^p k_a w_a~, &\qquad \sum_{b=1}^{a-1} w_b k_b = N - w_a k_a - \sum_{b=a+1}^p w_b k_b~,
\end{array}}
we can rewrite this as (see also \cite[(5.22), (6.20), (6.23)--(6.25)]{Bah:2022yjf})
\bes{\label{eq:arho}
\scalebox{0.9}{$
\begin{split}
a_\rho &= \frac{1}{48}N(8N^2-3) \\
& \qquad -\frac{1}{4} \sum_{a=1}^p \left(N k_a \left(w_a^2-\frac{1}{4}\right) -\frac{w_a^3 k_a^2}{3} + \sum_{b=a+1}^p k_a k_b \left(\frac{w_a^3}{3} - w_a^2 w_b -\frac{w_a-w_b}{4}\right)\right)~.
\end{split}$}
}

Let us demonstrate this formula with several examples. For a completely closed (trivial) puncture $[N]$, we have $p=1$, $k_1=1$, $\ell_1=1$, and $w_1=N$, which gives
\bes{
a_{[N]} = \frac{1}{48}N(8N^2-3) - \frac{1}{4} \left(\frac{2}{3}N^3-\frac{1}{4} N \right) = 0~,
}
as expected. For a minimal puncture $[N-1,1]$, we have $p=2$, $k_1=k_2=1$, $\ell_1=2$, $\ell_2=1$, $w_1=1$, and $w_2=N-1$. This yields
\bes{\label{eq:a-simple}
a_{[N-1,1]} = \frac{1}{24}(6 N^2-5)~;
}
the leading term in $N$ is $N^2/4$. As a generalization of this, we can consider the partition $[N-m,m]$. For this, we set $p=2$, $k_1=k_2=1$, $l_1=2$, $l_2=1$, $w_1=m$, and $w_2=N-m$. Then \eqref{eq:arho} gives
\begin{equation}\label{eq:a-small}
	a_{[N-m,m]} = \frac{m}{24}(6 N^2-8 m^2 +3)\,,
\end{equation}
Finally consider a maximal puncture $[1^N]$. Here we have $p=1$, $k_1=N$, $\ell_1=N$, and $w_1=1$, which gives
\bes{
a_{[1^N]} = \frac{1}{48}N(N-1)(8N+3)~,
}
with a leading term of $N^3/6$.

In this subsection we have reviewed anomalies for the ${\mathcal N}=2$ class S. As we mentioned in Sec.~\ref{sub:ads5}, the $F_0=0$ case of our AdS$_5$ solutions actually reproduces the  ${\mathcal N}=1$ MN solutions. As we reviewed in \eqref{eq:27/32}, the $a$ anomaly for these SCFTs is obtained by simply multiplying the ${\mathcal N}=2$ result \eqref{eq:arho} by a factor of $27/32$.

We will soon compare the puncture contribution in \eqref{eq:arho} to our earlier gravity results. For now, let us compare its first term, the bulk contribution, with the first term in \eqref{eq:a4-a6-intro}. For the 6d $\mathcal{N}=(2,0)$ $A_{N-1}$ theory, $a_6\sim \frac{16}7 N^3$ at large $N$; this gives
\begin{equation}
    \frac9{32}(g-1)N^3=\frac{27}{32}\frac13 (g-1)N^3
\end{equation}
This is indeed $27/32$ times the large $N$ behavior of the first term in \eqref{eq:arho}. 

\subsection{Match with gravity}
\label{sub:match}

We will now compare the puncture contribution in \eqref{eq:arho} with our gravity results. 

We begin with the minimal puncture \eqref{eq:a-simple}.  Recalling our comment below \eqref{eq:partN}, this is identified with a single defect M5, which would be expected to reduce to a D4 in IIA.\footnote{In this case, it is also possible to consider directly an M5 probe in the $\mathcal{N}=2$ MN background. Take the coordinate system in \cite[(2.7]{Gaiotto:2009gz}. From \cite[Sec.~4.2]{Bah:2013wda} we find that a BPS M5 is wrapped on $\{\theta=\theta_0, \, \psi=\pi/2,\, \phi=2\chi\}$, with $\theta_0 =\pi/2$ fixed by demanding that $U(2)_{\rm R}$ be preserved. The DBI term is in $T_{M5}16\pi^3 N^2 l_p^6 (1+\cos^2(\theta_0)$. The potential $A_6=-4\pi^2 N^2 l_p^6 (2\sin^2 \theta \dd \chi -4(3+\cos^2\theta)r^2 \dd\beta/(1-r^2)^2)$ also gives rise to a WZ term $T_{M5} 16\pi^3 N^2 l_p^6 \sin^2 \theta_0$. Summing the two terms and using \eqref{eq:a-brane} gives $a_{{\rm M}5}=N^2/4$.} From the D4 brane action \eqref{eq:SD4-5} and our formula \eqref{eq:a-brane} relating it to the defect $a$ anomaly, we obtained in \eqref{eq:aD4D6-0}:
\begin{equation}\label{eq:aD4-0}
    a_\mathrm{D4}= \frac{27}{32}k\frac{N^2}4\,.
\end{equation}
For class S, $k=1$; we will discuss the case of general $k$ in the next subsection. We also need to recall again the factor $27/32$ relating the $\mathcal{N}=1$ and $\mathcal{N}=2$ versions of class S theories from \eqref{eq:27/32}. With this, \eqref{eq:aD4-0} does match the leading behavior with $N$ of \eqref{eq:a-simple}.

More generally, we can consider the puncture $[N-m,m]$, which corresponds to $m$ defect M5s, on a $\mathds{Z}_{k=1}$ point. Upon reduction to IIA, one expects this to become a bound state of a D6-brane with $w=m$ D4-branes. In \eqref{eq:aD4D6-0} we found:
\begin{equation}\label{eq:aD6-0}
    a_{\mathrm{D}6/\mathrm{D4}}= \frac{27}{32}k\frac m4 \left(N^2-\frac43 m^2\right)\,.
\end{equation}
Again with $k=1$ and recalling the $27/32$ factor, this matches \eqref{eq:a-small} when $N$ and $m$ are $\gg 1$.

While the $m$ in $[N-m,m]$ represents the number of defect D4-branes in IIA puncture, and of M5-branes in M-theory, the $N-m$ does not: it provides the remaining flux so that the integral of $F_4$ along the internal sphere is $N$. This is perhaps clarified by thinking about the extreme case $m=0$, namely $[N]$, when the puncture is absent. A puncture of this type has $m<N-m$, or in other words $m<N/2$. This is indeed the possible range we got from \eqref{eq:D6D4-DBI-q-5}.
It is remarkable that we achieve a match for this whole range, even if when $m$ is comparable with $N$ the D6/D4 can no longer be considered a probe.

On the other hand, with the present methods it is not possible to achieve a match for punctures that are still more general than $[N-m,m]$. One might imagine, for example, that the probe approximation is still appropriate for $[N-m_1,m_1,\ldots,m_k]$ when $m_i \ll N$.  From \eqref{eq:arho}, the leading behavior for this is $\sim\frac14 \sum_i m_i N^2 + p_3(m_i)$, with $p_3$ a cubic polynomial. The first term is just like in \eqref{eq:a-small}, but $p_3$ contains terms $m_i^2 m_j$ that represent the effect of a brane on another, and thus cannot be captured by the probe approximation. Presumably a IIA counterpart of the approach in \cite{Bah:2019jts} would be necessary to reproduce such terms.

\subsection{\texorpdfstring{Class $\mathrm{S}_k$}{Class Sk}}
\label{sub:Sk}

We now consider the field theory realization of \eqref{eq:aD4D6-0} for a general integer $k$. It is natural to study 4d $\mathcal{N}=1$ SCFTs obtained by compactifying the six-dimensional $\mathcal{N}=(1,0)$ theory, which lives on $N$ M5-branes on a $\mathbb{C}^2/\mathbb{Z}_k$ orbifold singularity, on a Riemann surface. These SCFTs are known as theories of class $\mathrm{S}_k$ \cite{Gaiotto:2015usa} (see also \cite{Franco:2015jna, Hanany:2015pfa, Razamat:2016dpl, Bah:2017gph, Kim:2017toz, Razamat:2022gpm} and \cite{Lykken:1997gy, Uranga:1998vf, Oh:1999sk}). In this paper, we focus on a specific configuration: the $\mathbb{Z}_k$ orbifold of a class $\mathrm{S}$ theory corresponding to one maximal puncture, a collection of minimal punctures, and one arbitrary puncture. In the notation of \cite[Pages 19, 23]{Hanany:2015pfa}, this corresponds to the class $\mathrm{S}_k$ theory denoted by $\mathcal{L}^{(n_L, \sigma_L= +), (n_R, \sigma_R= +)}_{n-2, n-2} \equiv \mathcal{L}^{++}_{n-2, n-2}$, where $n-2$ is the number of minimal punctures and $n_{L}, n_R = 0,1, \ldots, k-1$ are the colours of the two maximal punctures, each with sign $\sigma_L=\sigma_R = +1$ and opposite orientation. The colours $n_L$ and $n_R$ of the maximal punctures must satisfy the condition $n_R-n_L = n-2$, as per \cite[(3.5)]{Hanany:2015pfa}.\footnote{Specifically, in \cite[(3.5)]{Hanany:2015pfa}, for the left hand side, we have $\sum_p \sigma_p^{(\min)} =n-2$, and for the right hand side, we have $\sigma_L^{(\max)} = \sigma_R^{(\max)} =1$, $o_L=-1$, and $o_R=+1$, and so the sum on the right hand side equals to $n_R-n_L$.}

A simple example is the $\mathbb{Z}_k$ orbifold of 4d $\mathcal{N}=2$ $\mathrm{SU}(N)$ SQCD with $2N$ flavours. The latter is realized on a sphere of type $A_{N-1}$ with two maximal and two minimal punctures, described by the quiver diagram:
\bes{ \label{SQCDquiv}
\begin{tikzpicture}[baseline, font=\scriptsize]
\node[gauge,label=below:{}] (SU4) at (0,0) {$N$};
\node[flavour,label=below:{}] (4L) at (-2,0) {$N$};
\node[flavour,label=below:{}] (4R) at (2,0) {$N$};
\draw (4L)--(SU4)--(4R);
\end{tikzpicture}
}
Here, the circular node denotes the $\mathrm{SU}(N)$ gauge group and each square node denotes $N$ fundamental hypermultiplets. The central charge of this theory is
\bes{
a_{\eqref{SQCDquiv}} = \frac{1}{48}(4 N^2) + \frac{1}{48} (N^2 - 1) + \frac{3}{16}(N^2 - 1)= \frac{1}{24} ( 7 N^2-5)~.
}
The $\mathbb{Z}_k$ orbifold of this theory, the $\mathcal{L}^{++}_{2, 2}$ theory, is described by the quiver:
\bes{ \label{orbifoldofSQCD}
\begin{tikzpicture}[baseline, font=\scriptsize]
\node[gauge,label=below:{}] (SU4a) at (0,1) {$N$};
\node[flavour,label=below:{}] (4La) at (-2,1) {$N$};
\node[flavour,label=below:{}] (4Ra) at (2,1) {$N$};
\node[gauge,label=below:{}] (SU4b) at (0,0)  {$N$};
\node[flavour,label=below:{}] (4Lb) at (-2,0) {$N$};
\node[flavour,label=below:{}] (4Rb) at (2,0) {$N$};
\node[gauge,label=below:{}] (SU4c) at (0,-1) {$N$};
\node[flavour,label=below:{}] (4Lc) at (-2,-1) {$N$};
\node[flavour,label=below:{}] (4Rc) at (2,-1) {$N$};
\draw[->] (4La) to (SU4a);
\draw[->]  (SU4a) to (4Ra);
\draw[->] (4Lb) to (SU4b);
\draw[->]  (SU4b) to (4Rb);
\draw[->] (4Lc) to (SU4c);
\draw[->]  (SU4c) to (4Rc);
\draw[->]  (SU4c) to (SU4b);
\draw[->]  (SU4b) to (SU4a);
\draw[->]  (SU4a) to (4Lb);
\draw[->]  (SU4b) to (4Lc);
\draw[->]  (4Ra) to (SU4b);
\draw[->]  (4Rb) to (SU4c);
\draw[->]  (SU4c) to (4La);
\draw[->]  (4Rc) to (SU4a);
\draw[->, bend right=30]  (SU4a) to (SU4c);
\end{tikzpicture}
}
The diagram consists of $k$ copies of \eqref{SQCDquiv} (we have taken $k=3$ for convenience). Each gauge node has 3 incoming and 3 outgoing arrows; in other words, the number of flavours for each $\mathrm{SU}(N)$ gauge node is $3N$. Consequently, the $R$-charge of each chiral field is $2/3$ (see also \cite[Page 29]{Hanany:2015pfa}). With $5k$ chiral multiplets in total, the central charge is
\bes{ \label{aorbifoldofSQCD}
a_{\eqref{orbifoldofSQCD}} = \frac{1}{48} N^2 (5k) + \frac{3}{16} (N^2 - 1)(k) = \frac{1}{48} k (14 N^2-9) = k a_{\eqref{SQCDquiv}}+\frac{k}{48}~.
}
The difference $a_{\eqref{orbifoldofSQCD}}-k a_{\eqref{SQCDquiv}}=k/48$ arises because the contribution of the traceless adjoint chiral multiplet in the $\mathcal{N}=2$ vector multiplet is $\frac{1}{48}(N^2-1)$, whereas the corresponding contribution in the $\mathcal{N}=1$ orbifolded theory is $\frac{1}{48} k N^2$. Nevertheless, in the large $N$ limit, this difference can be neglected, and the central charge of the $\mathbb{Z}_k$ orbifolded theory is $k$ times that of the original theory, as expected.

This analysis can be generalized by partially closing one of the maximal punctures. For class S, the corresponding partitions determine the quivers in a well-known fashion; see for example \cite[Table 1]{Tachikawa:2009rb}. For concreteness, we consider a sphere of type $A_{N-1}$ with one maximal puncture, $N-m+1$ minimal punctures, and one $[N-m, m]$ puncture. The corresponding 4d $\mathcal{N}=2$ theory is described by the quiver:
\begin{equation} \label{4dN2quivmaxminsmall}
\scalebox{0.65}{
\begin{tikzpicture}[
    node distance=0.8cm and 0.8cm,
    gauge/.style={draw, circle, minimum size=1.5cm, inner sep=1pt},
    flavor/.style={draw, regular polygon, regular polygon sides=4, minimum size=1.5cm, inner sep=1pt},
    every node/.style={font=\scriptsize} 
]
\node[flavor] (n1) {$N$};
\node[gauge, right=of n1] (n2) {$N$};
\node[flavor, below=of n2] (fx) {$1$};
\node[right=of n2] (dots1) {$\cdots$};
\node[gauge, right=of dots1] (n3) {$2m+2$};
\node[gauge, right=of n3] (n4) {$2m+1$};
\node[gauge, right=of n4] (n5) {$2m$};
\node[flavor, below=of n5] (f1) {$1$};
\node[gauge, right=of n5] (n6) {$2m-2$};
\node[right=of n6] (dots2) {$\cdots$};
\node[gauge, right=of dots2] (n7) {$4$};
\node[gauge, right=of n7] (n8) {$2$};

\draw[thick] (n1) -- (n2);
\draw[thick] (n2) -- (dots1);
\draw[thick] (dots1) -- (n3);
\draw[thick] (n3) -- (n4);
\draw[thick] (n4) -- (n5);
\draw[thick] (n5) -- (n6);
\draw[thick] (n6) -- (dots2);
\draw[thick] (dots2) -- (n7);
\draw[thick] (n7) -- (n8);

\draw[thick] (n5) -- (f1);
\draw[thick] (n2) -- (fx);
\end{tikzpicture}}
\end{equation}
In this diagram, the number of colours in each gauge node decreases by $1$ from $N$ to $2m$, and then decreases by $2$ from $2m$ to $2$. Each $\mathrm{SU}(N_c)$ gauge node has $2N_c$ fundamental hypermultiplets, and there are $N-m$ gauge nodes in total. The central charge of this theory is
\bes{ \label{centralcharge4dN2}
a_{\eqref{4dN2quivmaxminsmall}} &= a_{\text{bulk}} + (N-m+1) a_{[N-1,1]} + a_{[N-m,m]} \\
&= \frac{1}{12}N^3 + \frac{7}{48}N^2 - \frac{7}{48}N - \frac{1}{3}m(m^2-1)~,
}
where the bulk contribution is $a_{\text{bulk}} = -\frac{1}{24}(N-1) (8 N^2 +8 N +5)$, and $a_{[N-m,m]}$ is given by \eqref{eq:a-small}. The same result can also be obtained directly from the quiver diagram. The $\mathbb{Z}_k$ orbifold of this theory is
\begin{equation} \label{orbmaxminsmall}
\scalebox{0.65}{
\begin{tikzpicture}[
    gauge/.style={draw, circle, minimum size=1.5cm, inner sep=1pt, text centered},
    flavor/.style={draw, regular polygon, regular polygon sides=4, minimum size=1.5cm, inner sep=1pt, text centered},
    flavor_blue/.style={draw=blue, fill=blue!10, regular polygon, regular polygon sides=4, minimum size=1.5cm, inner sep=1pt, text centered},
    flavor_green/.style={draw=green!50!black, fill=green!10, regular polygon, regular polygon sides=4, minimum size=1.5cm, inner sep=1pt, text centered},
    every node/.style={font=\scriptsize}, 
    node distance=0.8cm and 0.8cm
]

\begin{scope}[yshift=3.5cm]
    \node[flavor] (n1a) {$N$};
    \node[gauge, right=of n1a] (n2a) {$N$};
    \node[right=of n2a] (dots1a) {$\cdots$};
    \node[gauge, right=of dots1a] (n3a) {$2m+2$};
    \node[gauge, right=of n3a] (n4a) {$2m+1$};
    \node[gauge, right=of n4a] (n5a) {$2m$};
    \node[gauge, right=of n5a] (n6a) {$2m-2$};
    \node[right=of n6a] (dots2a) {$\cdots$};
    \node[gauge, right=of dots2a] (n7a) {$4$};
    \node[gauge, right=of n7a] (n8a) {$2$};
    
    \node[flavor_blue, above right=of n2a, yshift=0.2cm, xshift=0.4cm] (f_extra_a) {$1$};
    \node[flavor_green, above right=of n5a, yshift=0.2cm, xshift=2.5cm] (g_extra_a) {$1$};

    \draw[->, thick] (n1a) -- (n2a);
    \draw[->, thick] (n2a) -- (dots1a);
    \draw[->, thick] (dots1a) -- (n3a);
    \draw[->, thick] (n3a) -- (n4a);
    \draw[->, thick] (n4a) -- (n5a);
    \draw[->, thick] (n5a) -- (n6a);
    \draw[->, thick] (n6a) -- (dots2a);
    \draw[->, thick] (dots2a) -- (n7a);
    \draw[->, thick] (n7a) -- (n8a);
    
\end{scope}

\begin{scope}[yshift=0cm]
    \node[flavor] (n1b) {$N$};
    \node[gauge, right=of n1b] (n2b) {$N$};
    \node[right=of n2b] (dots1b) {$\cdots$};
    \node[gauge, right=of dots1b] (n3b) {$2m+2$};
    \node[gauge, right=of n3b] (n4b) {$2m+1$};
    \node[gauge, right=of n4b] (n5b) {$2m$};
    \node[gauge, right=of n5b] (n6b) {$2m-2$};
    \node[right=of n6b] (dots2b) {$\cdots$};
    \node[gauge, right=of dots2b] (n7b) {$4$};
    \node[gauge, right=of n7b] (n8b) {$2$};
    
    \node[flavor_blue, above right=of n2b, yshift=0.2cm, xshift=0.4cm] (f_extra_b) {$1$};
    \node[flavor_green, above right=of n5b, yshift=0.2cm, xshift=2.5cm] (g_extra_b) {$1$};

    \draw[->, thick] (n1b) -- (n2b);
    \draw[->, thick] (n2b) -- (dots1b);
    \draw[->, thick] (dots1b) -- (n3b);
    \draw[->, thick] (n3b) -- (n4b);
    \draw[->, thick] (n4b) -- (n5b);
    \draw[->, thick] (n5b) -- (n6b);
    \draw[->, thick] (n6b) -- (dots2b);
    \draw[->, thick] (dots2b) -- (n7b);
    \draw[->, thick] (n7b) -- (n8b);

\end{scope}

\begin{scope}[yshift=-3.5cm]
    \node[flavor] (n1c) {$N$};
    \node[gauge, right=of n1c] (n2c) {$N$};
    \node[right=of n2c] (dots1c) {$\cdots$};
    \node[gauge, right=of dots1c] (n3c) {$2m+2$};
    \node[gauge, right=of n3c] (n4c) {$2m+1$};
    \node[gauge, right=of n4c] (n5c) {$2m$};
    \node[gauge, right=of n5c] (n6c) {$2m-2$};
    \node[right=of n6c] (dots2c) {$\cdots$};
    \node[gauge, right=of dots2c] (n7c) {$4$};
    \node[gauge, right=of n7c] (n8c) {$2$};

    \node[flavor_blue, above right=of n2c, yshift=0.2cm, xshift=0.4cm] (f_extra_c) {$1$};
    \node[flavor_green, above right=of n5c, yshift=0.2cm, xshift=2.5cm] (g_extra_c) {$1$};

    \draw[->, thick] (n1c) -- (n2c);
    \draw[->, thick] (n2c) -- (dots1c);
    \draw[->, thick] (dots1c) -- (n3c);
    \draw[->, thick] (n3c) -- (n4c);
    \draw[->, thick] (n4c) -- (n5c);
    \draw[->, thick] (n5c) -- (n6c);
    \draw[->, thick] (n6c) -- (dots2c);
    \draw[->, thick] (dots2c) -- (n7c);
    \draw[->, thick] (n7c) -- (n8c);
    
\end{scope}

\draw[->, thick] (n2c) -- (n2b);
\draw[->, thick] (n3c) -- (n3b);
\draw[->, thick] (n4c) -- (n4b);
\draw[->, thick] (n5c) -- (n5b);
\draw[->, thick] (n6c) -- (n6b);
\draw[->, thick] (n7c) -- (n7b);
\draw[->, thick] (n8c) -- (n8b);

\draw[->, thick] (n2b) -- (n2a);
\draw[->, thick] (n3b) -- (n3a);
\draw[->, thick] (n4b) -- (n4a);
\draw[->, thick] (n5b) -- (n5a);
\draw[->, thick] (n6b) -- (n6a);
\draw[->, thick] (n7b) -- (n7a);
\draw[->, thick] (n8b) -- (n8a);

\draw[->, thick] (n2a) -- (n1b);
\draw[->, thick] (dots1a) -- (n2b); 
\draw[->, thick] (n4a) -- (n3b);
\draw[->, thick] (n5a) -- (n4b);
\draw[->, thick] (n6a) -- (n5b);
\draw[->, thick] (dots2a) -- (n6b); 
\draw[->, thick] (n8a) -- (n7b);
\draw[->, thick] (n7a) to (dots2b);
\draw[->, thick] (n3a) to (dots1b); 
\draw[->, thick] (n7c) to (dots2a); 

\draw[->, thick] (n2b) -- (n1c);
\draw[->, thick] (dots1b) -- (n2c); 
\draw[->, thick] (n4b) -- (n3c);
\draw[->, thick] (n5b) -- (n4c);
\draw[->, thick] (n6b) -- (n5c);
\draw[->, thick] (dots2b) -- (n6c); 
\draw[->, thick] (n8b) -- (n7c);
\draw[->, thick] (n7b) to (dots2c);
\draw[->, thick] (n3b) to (dots1c); 

\draw[->, thick, bend right=20] (n2a) to (n2c);
\draw[->, thick, bend right=20] (n3a) to (n3c);
\draw[->, thick, bend right=20] (n4a) to (n4c);
\draw[->, thick, bend right=20] (n5a) to (n5c);
\draw[->, thick, bend right=20] (n6a) to (n6c);
\draw[->, thick, bend right=20] (n7a) to (n7c);
\draw[->, thick, bend right=20] (n8a) to (n8c);

\draw[->, thick] (n2c) to (n1a);
\draw[->, thick] (n4c) to (n3a);
\draw[->, thick] (n5c) to (n4a);
\draw[->, thick] (n6c) to (n5a);
\draw[->, thick] (n8c) to (n7a);

\draw[->, thick, blue] (n2a) to (f_extra_a);
\draw[->, thick, blue] (f_extra_a) to (n2b);
\draw[->, thick, blue] (n2b) to (f_extra_b);
\draw[->, thick, blue] (f_extra_b) to (n2c);
\draw[->, thick, blue] (n2c) to (f_extra_c);
\draw[->, thick, blue] (f_extra_c) to (n2a);

\draw[->, thick, color=green!50!black] (n5a) to (g_extra_a);
\draw[->, thick, color=green!50!black] (g_extra_a) to (n5b);
\draw[->, thick, color=green!50!black] (n5b) to (g_extra_b);
\draw[->, thick, color=green!50!black] (g_extra_b) to (n5c);
\draw[->, thick, color=green!50!black] (n5c) to (g_extra_c);
\draw[->, thick, color=green!50!black] (g_extra_c) to (n5a);

\end{tikzpicture}
}
\end{equation}
The diagram above depicts the case of $k=3$, but the general case is obtained by stacking $k$ copies of \eqref{4dN2quivmaxminsmall} and connecting them with the same arrow pattern. The blue and green colours are used to improve the diagram's readability. For each gauge node with label $N_c$, the sum of labels from nodes connected by incoming arrows is $3N_c$, which equals the sum for outgoing arrows. This guarantees that the theory is free of gauge anomalies and that the number of flavours for each $\mathrm{SU}(N_c)$ group is $3N_c$. It follows that the $R$-charge of each chiral field is $2/3$. The central charge is therefore
\bes{ \label{centralchargeorb}
a_{\eqref{orbmaxminsmall}} &= k \left[ \frac{1}{12} N^3 + \frac{7}{48} N^2 - \frac{1}{8} N - \frac{1}{48}m (16m^2-15) \right]\\
&= k a_{\eqref{4dN2quivmaxminsmall}} + \frac{k}{48}(N-m)~.
}
As before, the difference $a_{\eqref{orbmaxminsmall}}- k a_{\eqref{4dN2quivmaxminsmall}}= \frac{k}{48}(N-m)$, which is of order one in $N$ and $m$, is due to the tracelessness of the adjoint chiral multiplet in the vector multiplet of each gauge node in \eqref{4dN2quivmaxminsmall}. In the large $N$ limit, the central charge of the orbifolded theory is $k$ times that of the original theory, as expected. Furthermore, by comparing \eqref{centralcharge4dN2} and \eqref{centralchargeorb}, we find that in the large $N$ limit, the contribution to the central charge from the $[N-m,m]$ puncture (with $1 \ll m \ll N$) in the orbifolded theory is
\bes{
\frac{1}{4} k m\left(N^2-\frac{4}{3} m^2\right)~,
}
which agrees with \eqref{eq:aD4D6-0} divided by the factor of $27/32$ for general $N$, $m$, and $k$.

\subsection{\texorpdfstring{$\mathcal{N}=1$ punctures beyond class $S_k$}{N=1 punctures beyond class Sk}} 
\label{sub:beyond}

We have studied BPS probes (D4-branes and D6/D4 bound states) in AdS$_{5}$ and AdS$_7$ solutions of massive IIA SUGRA, which holographically correspond to $\mathcal{N}=(1,0)$ SCFTs. The probes in AdS$_5$ correspond to punctures in the orthogonal Riemann surface directions. The number of punctures and their data together with the reduction of the bulk 6d SCFT on the Riemann surface leads to the 4d SCFT. All punctures contribute to the conformal anomaly. Our probe computation in the solutions of massive IIA, exemplified in section \ref{sec:massivedefects}, provides a prediction for the conformal anomaly contribution of new type of $\mathcal{N}=1$ punctures. 

Let us give two examples. The solutions with two stacks of D8-branes and regular poles \eqref{eq:regD8sol} are dual to 6d SCFTs whose low-energy effective description in the tensor branch is provided by symmetric linear quivers with $SU(r_i)$ (with $i=1,\ldots N$) gauge groups, where the nodes of the quiver tails have descending group ranks: see \eqref{eq:quivregD8}. For these theories, our two types of punctures give contributions
\begin{equation}
    a_{\mathrm{D}4}= \frac{9}{128} k \left(3 N^2-4 \mu ^2\right) \, ,\qquad
	a_{\mathrm{D}6/\mathrm{D}4} = \frac{9}{128} k m \left(3 N^2-4 \left(\mu ^2+m^2\right)\right).
\end{equation}

Similarly, the solution with an O8-plane \eqref{eq:massEstr} holographically corresponds to the $\mathcal N=(1,0)$ 6d SCFT called massive E-string in \cite{Bah:2017wxp}, whose tensor branch quiver was given in \eqref{eq:quivmassEstr}. Our BPS brane probes predict the existence of punctures in this massive E-string theory, whose anomaly contribution is
\begin{equation}\label{eq:aD4D6-Estring}
    a_{\mathrm{D}4}= \frac{9 n_0 N^3}{32}  \, ,\qquad
	a_{\mathrm{D}6/\mathrm{D}4} = \frac{9}{256}  n_0 m \left(4 N^3- m^3\right)\,.
\end{equation}
Strictly speaking, in order for the probe limit to hold we need $m\ll N/2$; recall however that in Sec.~\ref{sub:match} we managed to obtain a match with known class S results for all $m< N/2$. Actually, the D4 puncture already appeared in  \cite{Bah:2017wxp} beyond the probe approximation: an AdS$_5$ solution exists with many punctures of this type, smearead all over $\Sigma$ but with back-reaction taken into account. The holographic $a$ anomaly computed there was indeed of the form \eqref{eq:a4-a6-intro}, with the puncture contribution matching thta in \eqref{eq:aD4D6-Estring}. 

It would be interesting to have a field theoretic understanding of these punctures in the spirit of \cite{Heckman:2016xdl}, as well as a field theoretic computation of the anomaly contributions of these new punctures. 

\acknowledgments
We would like to thank F.~Bonetti, M.~De Marco, S.~Razamat, G.~Zafrir for interesting discussions. 
This work is supported in part by the INFN. The work of F.A.~is supported in part by the Italian MUR Departments of Excellence grant 2023-2027 ``Quantum Frontiers'' and by the University of Padua under the 2023 STARS Grants@Unipd programme (GENSYMSTR – Generalized Symmetries from Strings and Branes), and it has been also partially supported by the grant NSF PHY-2309135 to the Kavli Institute for Theoretical Physics (KITP). N.M.~gratefully acknowledges support from the Simons Center for Geometry and Physics, Stony Brook University, at which some of the research for this paper was performed during the 22nd Simons Physics Summer Workshop 2025. N.M.'s research is partially supported by the MUR-PRIN grant No. 2022NY2MXY (Finanziato dall'Unione europea -- Next Generation EU, Missione 4 Componente 1 CUP H53D23001080006, I53D23001330006). N.M.~also would like to thank to Fr\'ed\'eric A. Dreyer and Aroonroj Mekareeya for their warm hospitality during the completion of this project. A.T.'s research is partially supported by the MUR-PRIN grant No. 2022YZ5BA2.

\bibliographystyle{JHEP}
\bibliography{DefectSCE}

\end{document}